%% file: cheating_switcher_paper_merged_arxiv.tex
\documentclass[11pt, letterpaper, onecolumn, peerreview]{IEEEtran}
\usepackage{epsfig}
\usepackage{graphicx}
\usepackage{amsmath}
\usepackage{amssymb}
\newtheorem{theorem}{\bf Theorem}[section]
\newtheorem{lemma}{\bf Lemma}[section]

\newtheorem{corollary}{\bf Corollary}[section]
\def\mX{\mathcal{X}}

\def\mF{\mathcal{F}}
\def\mP{\mathcal{P}}
\def\mL{\mathcal{L}}
\def\mQ{\mathcal{Q}}
\def\mG{\mathcal{G}}
\def\mC{\mathcal{C}}
\def\mD{\mathcal{D}}
\def\mW{\mathcal{W}}

\def\mB{\mathcal{B}}

\def\mT{\mathcal{T}}
\def\mV{\mathcal{V}}
\def\mO{\mathcal{O}}
\def\conv{\textrm{\bf{conv}}}

\def\mXh{\widehat{\mathcal{X}}}
\def\x{\mathbf{x}}
\def\hx{\widehat{x}}
\def\xh{\widehat{\bf x}}
\def\wt{\widetilde}
\def\E{\mathbb{E}}
\def\R{\mathbb{R}}
\def\Z{\mathbb{Z}}
\def\s{\mathbf{s}}
\def\t{\mathbf{t}}
\def\1{\mathbf{1}}

\def\B{\bigg}
\def\h{\Bigg}
\def\l{\left}
\def\r{\right}

\def\wt{\widetilde}
\def\eps{\epsilon}

\def\be{\begin{equation}}
\def\ee{\end{equation}}
\def\ba{\begin{eqnarray}}
\def\ea{\end{eqnarray}}

\begin{document}
\title{The source coding game with a cheating switcher}
\author{Hari Palaiyanur, \IEEEmembership{Student Member}, Cheng Chang, Anant Sahai, \IEEEmembership{Member}
\thanks{This work was supported by an NSF Graduate Research Fellowship. An earlier version \cite{ISIT07} of the material in this paper was presented in part at the 2007 International Symposium on
Information Theory, Nice, France.}
\thanks{
The authors are with Wireless Foundations, Department of Electrical Engineering and Computer Sciences at the University of California at Berkeley. (email: hpalaiya@eecs.berkeley.edu;
cchang@eecs.berkeley.edu; sahai@eecs.berkeley.edu)}} \maketitle

\begin{abstract}
 Motivated by the lossy compression of an active-vision video stream, we consider the problem of
 finding the rate-distortion function of an arbitrarily varying
 source (AVS) composed of a finite number of subsources with known distributions. Berger's paper `The Source Coding Game',
 \emph{IEEE Trans. Inform. Theory}, 1971, solves this problem under the condition that the adversary is allowed
 only strictly causal access to the subsource realizations. We consider the case when the adversary has
 access to the subsource realizations non-causally. Using the type-covering lemma, this new rate-distortion
 function is determined to be the maximum of the IID rate-distortion function over a
 set of source distributions attainable by the adversary. We then extend the results to allow for
 partial or noisy observations of subsource realizations. We further explore the model by
 attempting to find the rate-distortion function when the adversary is actually helpful.

 Finally, a bound is developed on the uniform continuity of the IID
 rate-distortion function for finite-alphabet sources. The bound is
 used to give a sufficient number of distributions that need to be
 sampled to compute the rate-distortion function of an AVS to within a
 certain accuracy. The bound is also used to give a rate of
 convergence for the estimate of the rate-distortion function for an
 unknown IID finite-alphabet source .
\end{abstract}


\section{Introduction} \label{sec:intro}

\subsection{Motivation}

    Active vision/sensing/perception \cite{BajcsyVision} is an approach to computer vision, the
main principle of which is that sensors should choose to explore their environment actively
{\em based on what they currently sense or have previously sensed}. As
Bajcsy states it in \cite{BajcsyVision}, ``We do not just see, we look." The contrast to passive
sensors can be seen by comparing a fixed security camera (non-active) to a person holding a
camera (active). Even if the person is otherwise stationary, they may zoom the camera
into any part of their visual field to obtain a better view (e.g. if they see a trespasser).  There is also the
possibility that the sensor has noncausal information about the environment. For example, a
cameraman at a sporting event generally has only causal knowledge of the environment. A
cameraman on a movie set, however, has noncausal information about the environment through the
script. The noncausal information can be advantageous to the cameraman in (actively) capturing
the important features of a scene.

    There is a subtle distinction between causal and strictly causal information and this
distinction is related to the time-scales on which the environment changes. A causal active
sensor knows both the present and the past, but a strictly causal one knows only the past. If
the environment changes at a pace much slower than the sensor can actively look, there is
essentially no difference between knowing the immediate past and knowing the present. However,
if the environment changes at a pace faster then the sensor can actively look (and process
information), there is intuitively a substantial difference between knowing only the past and
knowing the present.

    As motivation for this paper, we are interested in the fixed-rate lossy compression of an active-vision
source. In reality, there are many interesting questions that need to be answered to truly
understand the problem, including:
\begin{itemize}
 \item What is the relevant distortion measure for active-video?
 \item Is there a distinction between the compression of an active-video source for use by the closed-loop control system that points the camera as compared to compression for later off-line use?
 \item How to model the entire plenoptic function that the active-video source will be dynamically sampling? \cite{VetterliPlenoptic}
\end{itemize}

It is also clear that the core issues here extend well beyond vision. They also arise in a
series of sensor-measurements that were dynamically sampled by a distributed sensor network as
well as the case of measurements taken by an autonomously moving sensor that chooses where to
go in part based on what it is observing. More provocatively, similar issues of active-sources
arise when the successive source symbols are brought by customers, each of which has free will
and can choose among competing codecs for compression.\footnote{This is related to a
particularly odd kind of
  moral hazard in private health insurance markets. Somewhat
  counterintuitively, private health insurers actually have a disincentive to
  provide good treatment of chronic conditions since they fear attracting
  patients that are intrinsically likely to get sick! \cite{longmanWashingtonMonthly}}

We concentrate entirely on the simplest aspect of the problem: what is
the impact on the rate-distortion function of having the source being
actively sampled by an entity that knows something about the
realizations of the environment as it does the sampling. Thus, we
assume an overly simplified traditional rate-distortion setting with
known finite alphabets and bounded distortion measures. The goal is
the traditional block-coding one: meet an average distortion
constraint with high probability using as little rate as possible.

The modeling question is whether or not it is worth building a
detailed model for how the active-source is going to be doing its
dynamic sampling of the source.  Three basic ways to model the goals of
the camera are worst case (adversarial), random (agnostic), and
helpful (joint optimization of camera and coding system). Admittedly,
the most interesting problems involve the compression of sources with
memory, but following tradition we focus on memoryless sources to
understand the basic differences between active and non-active sources
for lossy compression.

In the context of active-vision, a strictly causal adversary pointing
a camera is intuitively no more threatening than a robot randomly
pointing the camera when the scene being captured is memoryless.  This
intuition was formally proved correct in \cite{BergerSourceCodingGame}
by Berger as he determined the rate-distortion function for memoryless
sources and a strictly causal adversarial model. This paper determines
the rate-distortion function for the additional cases of causal and
non-causal adversaries. The model is then extended to allow only noisy
observations by the adversary doing the sampling of the scene. To see
the impact of the details of the dynamic sampling on the
rate-distortion function, the paper also considers how the
rate-distortion function changes when the `adversary' is actually a
helpful party.

\subsection{Causality in information theory}

    The issue of causality arises naturally in several major problems of information theory where
noncausal knowledge of the realizations of randomness in the problem can be advantageous.
Shannon \cite{ShannonCausalSI} studied the problem of transmitting information over a noisy
channel with memoryless state parameter revealed to the encoder causally. Gelfand and Pinsker
\cite{GelfandPinsker} studied the same problem with the state parameter available to the
encoder noncausally. In general, the capacity is larger when the channel state is available
noncausally to the encoder. When the channel state corresponds to Gaussian interference known
noncausally, Costa \cite{CostaDirtyPaperCoding} showed that the capacity is the same as when
the interference is not present at all. Willems (\cite{WillemsDirtyTape1},
\cite{WillemsDirtyTape2}) gave achievable strategies when the Gaussian interference is known
only causally. Lattice strategies for both causal and non-causal knowledge of the interference
are discussed in \cite{ErezLatticesInterference}, but the advantage of finitely anticipatory
knowledge of interference is not yet explicitly understood even in the case of Gaussian
interference.

    Agarwal et.al. \cite{Agarwal} find the capacity for an arbitrarily varying channel whose
input is constrained to look like an IID source with known distribution. The adversary is
constrained to distort over a block to at most some (additive) distortion, but is not
constrained to act causally. \cite{Agarwal} shows that the rate-distortion function turns out
to be the capacity for this channel. Because the codewords are
constrained to look IID, simulating the action of a causal memoryless
channel turns out to be sufficient for the adversary to minimize the capacity.

Causality also has implications for the problem of lossy source
coding, as studied by Neuhoff and Gilbert
\cite{NeuhoffGilbert}. There, for an IID source, causal source codes
generally require a higher rate to achieve distortion $D$ than
non-causal source codes. It is also shown that optimal causal source
codes can be constructed by time-sharing between memoryless codes.
Hence, there is a rate penalty for using causal coders (as opposed to
noncausal coders), but no further penalty for using memoryless
coders. Similar results have been derived by Weissman and Merhav
\cite{WeissmanMerhavSI} for lossy source coding with causal and
noncausal side information. In \cite{NeuhoffGilbert}, the channel was
implicitly assumed to noiseless and binary. Tatikonda, et.al
\cite{TatikondaSRD} show that even if the channel is matched properly
to achieve the {\em sequential} rate-distortion function, there is a
penalty for using causal coders when the sources have memory. For
example, they show that proper matching for a Gauss-Markov source is a
Gaussian channel with feedback, but the rate-distortion performance
with this causal matching still does not meet the performance of
noncausal coders.

\subsection{Results and organization of paper}
    Section \ref{sec:setup} sets up the notation, model and briefly reviews the literature on
lossy compression of arbitrarily varying sources. Section \ref{sec:result} gives the
rate-distortion function for an AVS when the adversary has noncausal access to realizations of
a finite collection of memoryless subsources and can sample among them. As shown in Theorem
\ref{thm:mainresult}, the rate-distortion
function for this problem is the maximization of the IID rate-distortion function over the
memoryless distributions the adversary can simulate. The adversary requires only causal information to
impose this rate-distortion function. This establishes that when the subsources are memoryless, the
rate-distortion function can strictly increase when the adversary has
knowledge of the present subsource realizations,
but no further increase occurs when the adversary is allowed knowledge
of the future.

We then extend the AVS model to include noisy or partial observations
of the subsource realizations and determine the rate-distortion
function for this setting in Section \ref{sec:states}. As shown in
Theorem \ref{thm:states}, the form of the solution is the same as for
the adversary with clean observations, with the set of attainable
distributions essentially being related to the original distributions
through Bayes' rule.

Next, Section \ref{sec:helpful} explores the problem when the goal of the active sensor is to
help the coding system achieve a low distortion. Theorem \ref{thm:helpfulcheating} gives a
characterization of the rate-distortion functions if the helper is fully noncausal in terms of
the rate-distortion function for an associated lossy compression problem. As a corollary, we
also give bounds for the cases of causal observations and noisy observations.

Simple examples illustrating these results are given in Section \ref{sec:examples}. In Section
\ref{sec:computingrd}, we discuss how to compute the rate-distortion function for arbitrarily
varying sources to within a given accuracy using the uniform continuity of the IID
rate-distortion function. The main tool there is an explicit bound on the uniform continuity of
the IID rate-distortion function that is of potentially independent interest. Finally, we
conclude in Section \ref{sec:conclusion}.

All the problems in this paper are studied in the context of
fixed-length block coding. Variable-length coding could perform better
in a universal sense by using only as much rate as required when the
active sensor is not adversarial. However, we are interested in
determining upper and lower bounds for the rate that active sensors
might end up needing and for this purpose, fixed-length block coding
is appropriate.

\section{Problem Setup} \label{sec:setup}
\subsection{Notation}

Let $\mX$ and $\mXh$ be the finite source and reconstruction alphabets respectively. Let $\x^n
= (x_1, \ldots, x_n)$ denote an arbitrary vector from $\mX^n$ and $\xh^n = (\hx_1, \ldots,
\hx_n)$ an arbitrary vector from $\mXh^n$. When needed, $\x^k = (x_1, \ldots, x_k)$ will be
used to denote the first $k$ symbols in the vector $\x^n$.

Let $d:\mX \times \mXh \rightarrow [0,d^\ast]$ be a distortion measure on the product set $\mX
\times \mXh$ with maximum distortion $d^\ast < \infty$. Let
    \begin{equation}   \widetilde{d} =  \min_{(x,\hx):~d(x,\hx) > 0} d(x,\hx)
        \label{eqn:mindistortion} \end{equation}
be the minimum nonzero distortion. Define $d_n:\mX^n \times \mXh^n \rightarrow [0,d^\ast]$ for $n\geq 1$ to be
    \begin{equation} d_n(\x^n, \xh^n) = \frac{1}{n} \sum_{k=1}^n d(x_k,\hx_k). \end{equation}

Let $\mP(\mX)$ be the set of probability distributions on $\mX$, let $\mP_n(\mX)$ be the set of types
of length $n$ strings from $\mX$, and let $\mW$ be the set of probability transition matrices
from $\mX$ to $\mXh$. Let $p_{\x^n} \in \mP_n(\mX)$ be the empirical type of a vector $\x^n$.
For a $p \in \mP(\mX)$, let
    \begin{equation} D_{\min}(p) = \sum_{x\in \mX} p(x) \min_{\hx \in \mXh} d(x,\hx) \end{equation}
be the minimum average distortion achievable for the source distribution $p$. The rate-distortion function of $p \in \mP(\mX)$ at distortion $D > D_{\min}(p)$ with respect
to distortion measure $d$ is defined to be
    \begin{equation}
        R(p,D) = \min_{W \in \mW(p,D)} I(p,W),
    \end{equation}
where
    \begin{equation}
        \mW(p,D) = \B \{ W \in \mW: \sum_{x \in \mX} \sum_{\hx \in \mXh} p(x)W(\hx|x)d(x,\hx) \leq D
        \B \}
    \end{equation}
and $I(p,W)$ is the mutual information\footnote{We use natural log, denoted $\ln$, and nats in
most of the paper. In examples only, we use bits. }
    \begin{equation}
        I(p,W) = \sum_{x \in \mX} \sum_{\hx\in \mXh} p(x) W(\hx|x)\ln \h[ \frac{W(\hx|x)}{\sum_{x' \in \mX}
         p(x')W(\hx|x')}\h].
    \end{equation}

Let $\mB = \{\xh^n(1), \ldots, \xh^n(K)\}$ be a codebook with $K$ length-$n$ vectors from
$\mXh^n$. Define
    \begin{equation}
        d_n(\x^n;\mB) = \min_{\xh^n \in \mB} d_n(\x^n, \xh^n).
    \end{equation}

If $\mB$ is used to represent an IID source with distribution $p$, then the average distortion
of $\mB$ is defined to be
    \begin{equation}
        d(\mB) = \sum_{\x^n \in \mX^n} P(\x^n)d_n(\x^n;\mB) = \E[d_n(\x^n;\mB)],
    \end{equation}
where
    \begin{equation}
        P(\x^n) = \prod_{k=1}^n p(x_k).
    \end{equation}
For $n \geq 1$, $D > D_{\min}(p)$, let $K(n,D)$ be the minimum number of codewords needed in a
codebook $\mB \subset \mXh^n$ so that $d(\mB) \leq D$. By convention, if no such codebook
exists, $K(n,D) = \infty$. Let the rate-distortion function\footnote{We define $R(D_{\min}(p))
= \lim_{D \downarrow D_{\min}(p)} R(D)$. This is equivalent to saying that a sequence of codes
represent a source to within distortion $D$ if their average distortion is tending to $D$ in
the limit. The only distortion where this distinction is meaningful is $D_{\min}(p)$.} of an
IID source be $R(D) = \limsup_n \frac{1}{n} \ln K(n,D)$. Shannon's rate-distortion theorem
(\cite{ShannonRateDistortion}, \cite{WolfowitzRateDistortion}) states that for all $n$,
$\frac{1}{n}\ln K(n,D) \geq R(p,D)$ and
    \begin{equation} \liminf_{n \to \infty} \frac{1}{n}\ln K(n,D) = R(D) = R(p,D).
    \end{equation}

\subsection{Arbitrarily varying sources} \label{sec:game}
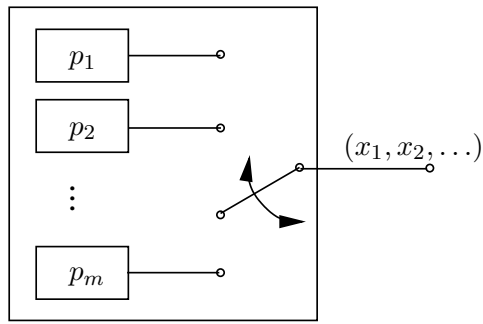
\begin{figure}
        \begin{center}
        \input{switcher.pstex_t}
        \end{center} \caption{A class of models for an AVS. The switcher can set the switch position according to the rules of the
        model.} \label{fig:switcher}
\end{figure}

The source coding game is a two-player game introduced in \cite{BergerSourceCodingGame} by
Berger as a model for an AVS. The two players are called the `switcher' and `coder'. In a
coding context, the coder corresponds to the designer of a lossy source code and the switcher
corresponds to a potentially malicious adversary pointing the camera.

Figure \ref{fig:switcher} shows a model of an AVS. There are $m$ IID `subsources' with common
alphabet $\mX$. In \cite{BergerSourceCodingGame}, the subsources are assumed to be independent,
but that restriction turns out not to be required\footnote{In \cite{BergerSourceCodingGame}, the motivation
was multiplexing data streams and independence is a reasonable assumption, but the proof does
not require it. Active sources, however, would likely choose among correlated subsources in practice.}.
There can be multiple subsources governed by the same distribution. In that sense, the switcher has access to
a {\em list} of $m$ subsources, rather than a set of $m$ different distributions. The marginal
distributions of the $m$ subsources are known to be $\{p_l\}_{l=1}^m$ and we let $\mG = \{p_1,
\ldots, p_m\}$. Let $P(x_{1,1}, \ldots, x_{m,1})$ be the joint probability distribution for the
IID source $\{(x_{1,k},\ldots,x_{m,k})\}_k$. Fix an $n \geq 1$ and consider a block of length
$n$. We let $x_{l,k}$ denote the output of the $l^{th}$ subsource at time $k$. We will use
$\x_l^n$ to denote the vector $(x_{l,1}, \ldots, x_{l,n})$. At each time $k$, the AVS outputs a
letter $x_k$ which is determined by the position of the switch inside the AVS. The switch
positions are denoted $\s^n = (s_1, \ldots, s_n)$ with $s_k \in \{1, 2, \ldots, m\}$ for each
$1 \leq k \leq n$. With this notation, $x_k = x_{s_k, k}$ for $1 \leq k \leq n$.

The switcher can set the switch position according to the model for the AVS. For example, in the
compound source setting of Sakrison \cite{SakrisonCompoundSources}, the switcher chooses $s \in
\{1, \ldots, m\}$ and sets $s_k = s$ for $1 \leq k \leq n$. The main case analyzed in
\cite{BergerSourceCodingGame} allowed the switcher to change $s_k$ arbitrarily, but the
switcher only had knowledge at time $k$ of $\s^{k-1}$ and $\x^{k-1}$. That is, the switcher
only had knowledge of past switch positions and past AVS outputs before deciding the switch
position at each time. One of the cases analyzed in this paper is termed full-lookahead, where
the switcher makes a (possibly random) decision about the full $\s^n$ with knowledge of $\x_1^n, \x_2^n,
\ldots, \x_m^n$ beforehand. The other case is termed $1$-step lookahead\footnote{We use the
term $1$-step lookahead even though this term is meant to represent the causal (but not
strictly) switcher. In most of the information theory literature, `causal' knowledge includes
knowledge of the present.}, where for each $k$, $s_k$ is a (possibly random) function of
$\x_1^k, \ldots, \x_m^k$. The switcher may or may not have knowledge of the codebook, but this
knowledge turns out to be inconsequential for the rate-distortion function.

The coder's goal is to design a codebook $\mB$ of minimal size to represent $\x^n$ to within
distortion $D$ on average. The codebook must be able to do this for {\em every} allowable
strategy for the switcher according to the model. Define
\begin{equation}
    M(n,D) = \min\left\{ |\mB| : \begin{array}{c}
        \mB \subset \mXh^n, ~ \E[d_n(\x^n;\mB)] \leq D \\
         \textrm{for all allowable} \\
            \textrm{switcher strategies }
        \end{array} \right\}.
\end{equation}

Here, $\E[d_n(\x^n; \mB)]$ is defined to be $\sum_{\x^n} \left ( \sum_{\s^n} P(\s^n,\x^n)
\right ) d_n(\x^n; \mB)$, where $P(\s^n,\x^n)$ is an appropriate probability mass function on
$\{1,\ldots,m\}^n \times \mX^n$ that agrees with the model of the AVS. When the switcher has
full lookahead, $P(\s^n, \x^n)$ must be composed of conditional distributions of the form
    \begin{equation}   P(\s^n,\x^n|\x_1^n, \ldots, \x_m^n) = P(\s^n|\x_1^n,\ldots, \x_m^n)\cdot \prod_{k=1}^n 1(x_k = x_{s_k,k}). \end{equation}
Then, $P(\s^n,\x^n)$ is simply obtained by averaging over $(\x_1^n,\ldots, \x_m^n)$.
    \begin{equation} P(\s^n,\x^n) = \sum_{(\x_1^n,\ldots, \x_m^n)} \l(\prod_{k=1}^n P(x_{1,k},\ldots,x_{m,k}) \r) P(\s^n|\x_1^n,\ldots,\x_m^n). \end{equation}
For a set of distributions $\mQ \subset \mP(\mX)$, let $D_{\min}(\mQ) = \sup_{p \in \mQ}
D_{\min}(p)$. We are interested in the exponential rate of growth of $M(n,D)$ with $n$. Define
the rate-distortion function of an AVS to be
   \begin{equation} R(D) \triangleq \limsup_{n \to \infty} \frac{1}{n} \ln M(n,D). \end{equation}
In every case considered, it will be also be clear that $R(D) = \liminf_{n\to \infty}
\frac{1}{n} \ln M(n,D)$.

\subsection{Literature Review}
\paragraph{One IID source}

Suppose $m = 1$. Then there is only one IID subsource $p_1 = p$ and the switch position is
determined to be $s_k = 1$ for all time. This is exactly the classical rate-distortion problem
considered by Shannon \cite{ShannonRateDistortion}, and he showed
    \begin{equation}
        R(D) = R(p,D).
    \end{equation}
Computing $R(p,D)$ can be done with the Blahut-Arimoto algorithm \cite{CsiszarBook}, and also
falls under the umbrella of convex programming.

\paragraph{Compound source}

Now suppose that $m > 1$, but the switcher is constrained to choose $s_k = s \in \{1, \ldots,
m\}$ for all $k$. That is, the switch position is set once and remains constant afterwards.
Sakrison \cite{SakrisonCompoundSources} studied the rate-distortion function for this class of
{\em compound} sources and showed that planning for the worst case subsource is both necessary
and sufficient. Hence, for compound sources,
    \begin{equation}
        R(D) = \max_{p \in \mG} R(p,D).
    \end{equation}
This result holds whether the switch position is chosen with or without knowledge of the
realizations of the $m$ subsources. Here, $R(D)$ can be computed easily since $m$ is finite and
each individual $R(p,D)$ can be computed.

\paragraph{Causal adversarial source} In Berger's setup \cite{BergerSourceCodingGame}, the switcher is allowed to choose $s_k \in
\{1,\ldots, m\}$ arbitrarily at any time $k$ , but must do so in a strictly causal manner
without access to the current time step's subsource realizations. More specifically, the switch
position $s_k$ is chosen as a (possibly random) function of $(s_1,\ldots, s_{k-1})$ and $(x_1,
\ldots,x_{k-1})$. The conclusion of \cite{BergerSourceCodingGame} is that under these rules,
    \begin{equation}
        R(D) = \max_{p \in \conv ( \mG)} R(p,D), \label{eqn:bergerscgamerd}
    \end{equation}
where $\conv ( \mG)$ is the convex hull of $\mG$. It should be noted that this same
rate-distortion function applies in the following cases:
    \begin{itemize}
        \item The switcher chooses $s_k$ at each time $k$ without {\em any} observations at all.
        \item The switcher chooses $s_k$ as a function of the first $k-1$ outputs of {\em all}
$m$ subsources.
    \end{itemize}
Note that in (\ref{eqn:bergerscgamerd}), evaluating $R(D)$ involves a maximization over an infinite
set, so the computation of $R(D)$ is not trivial since $R(p,D)$ is not necessarily a concave $\cap$
function. A simple, provable, approximate (to any given accuracy) solution is discussed in
Section \ref{sec:computingrd}.

\section{$R(D)$ for the cheating switcher} \label{sec:result}
In the conclusion of \cite{BergerSourceCodingGame}, Berger poses the question of what happens
to the rate-distortion function when the rules are tilted in favor of the switcher. Suppose
that the switcher were given access to the $m$ subsource realizations before having to choose
the switch positions; we call such a switcher a `cheating switcher'. In this paper, we deal
with two levels of noncausality and show they are essentially the same when the subsources are
IID over time:
    \begin{itemize}
        \item The switcher chooses $s_k$ based on the realizations of the $m$ subsources at
        time $k$. We refer to this case as $1$-step lookahead for the switcher.
        \item The switcher chooses $(s_1, \ldots, s_n)$ based on the entire length $n$ realizations of
        the $m$ subsources. We refer to this case as full lookahead for the switcher.
    \end{itemize}

\begin{theorem}\label{thm:mainresult}
Suppose the switcher has $1$-step lookahead or full lookahead. In both cases, for $D >
D_{\min}(\mC)$,
\begin{equation} R(D) = \widetilde{R}(D) \triangleq \max_{p \in \mC} R(p,D), \label{eqn:cheatingswitcherrd} \end{equation}
where

\begin{equation} \mC = \left\{ \begin{array}{ccc}
& & \sum_{i\in \mathcal{V}} p(i) \geq P\big(x_l \in \mathcal{V}, 1 \leq l \leq m \big) \\
   p \in \mP & : & \forall~ \mathcal{V} \textrm{ such that } \\
 &  &  \mathcal{V} \subseteq \mX
\end{array} \right\}. \label{eqn:defnC}\end{equation}

For $D < D_{\min}(\mC)$, $R(D) = \infty$ by convention because the switcher can simulate a
distribution for which the distortion $D$ is infeasible for the coder.
\end{theorem}
{\em Remarks:}
\begin{itemize}
\item If there are at least two non-deterministic subsources and $\conv(\mG) \neq \mP(\mX)$,
then $\conv(\mG)$ is a strict subset of $\mC$, and thus $R(D)$ can strictly increase when the switcher is
allowed to look at the present subsource realizations before choosing the switch position. Hence,
extra rate must be provisioned for active sensors in general.

\item As a consequence of the theorem, we see that when the subsources within an AVS are IID,
knowledge of past subsource realizations is useless to the switcher, knowledge of the current
step's subsource realizations is useful, and knowledge of future subsource realizations beyond
the current step is useless if $1$-step lookahead is already given.

\item Note that computing $R(D)$ requires further discussion given in Section
\ref{sec:computingrd}, just as it does for the strictly causal case of Berger.
\end{itemize}

\begin{proof} We give a short outline of the proof here. See Appendix \ref{sec:appendix1} for the complete proof.
To show $R(D) \leq \wt{R}(D)$, we use the type-covering lemma from
\cite{BergerSourceCodingGame}. It says for a fixed type $p$ in $\mP_n(\mX)$ and $\eps > 0$, all
sequences with type $p$ can be covered within distortion $D$ with at most $\exp(n(R(p,D) +
\eps))$ codewords for large enough $n$. Since there are at most $(n+1)^{|\mX|}$ distinct types, we can
cover all $n$-length strings with types in $\mC$ with at most $\exp(n(\wt{R}(D)+
\frac{|\mX|}{n}\ln(n+1) + \eps))$ codewords. Furthermore, we can show that types not in $\mC$
occur exponentially rarely even if the switcher has full lookahead, meaning that their
contribution to the average distortion can be bounded by $d^\ast$ times an exponentially
decaying term in $n$. Hence, the rate needed regardless of the switcher strategy is at most
$\wt{R}(D) + \eps$ with $\eps > 0$ arbitrarily small.

    Now, to show $R(D) \geq \wt{R}(D)$, we describe one potential strategy for the adversary.
This strategy requires only $1$-step lookahead and it forces the coder to use rate at least
$\wt{R}(D)$.  For each
set $\mV \subset \mX$ with $\mV \neq \emptyset$ and $|\mV| \leq m$, the adversary has a random
rule $f(\cdot|\mV)$, which is a probability mass function (PMF) on $\mV$. At each time $k$, if
the switcher observes a candidate set $\{x_{1,k}, \ldots, x_{m,k}\}$, the switcher chooses to
output $x \in \{x_{1,k}, \ldots, x_{m,k}\}$ with probability $f(x|\{x_{1,k}, \ldots,
x_{m,k}\})$. If $\beta(\mV) = P(\{x_{1,k},\ldots, x_{m,k}\} = \mV)$, let
    \begin{equation}
        \mD \triangleq \left\{ \begin{array}{ccc}
        & & p(x) = \sum_{\mathcal{V} \subseteq \mX, |\mathcal{V}|\leq m} \beta(\mathcal{V}) f(x|\mathcal{V}), x \in \mX\\
         p \in \mP & : & f(\cdot|\mathcal{V}) \textrm{ is a PMF on} ~\mathcal{V}, \\
        &  & \forall~ \mathcal{V} \textrm{ s.t. } \mathcal{V} \subseteq \mX, ~ |\mathcal{V}| \leq m  \end{array}
        \right\}.
    \label{eqn:alternativeset}\end{equation}
$\mD$ is the set of IID distributions the AVS can `simulate' using these memoryless rules
requiring $1$-step lookahead. It is clear by construction that $\mD \subseteq \mC$. Also, it is
clear that both $\mC$ and $\mD$ are convex sets of distributions. Lemma \ref{lem:hyperplanearg}
in Appendix \ref{sec:appendix1} uses a separating hyperplane argument to show $\mD = \mC$. The
adversary can therefore simulate any IID source with distribution in $\mC$ and hence $R(D) \geq
\wt{R}(D)$.
\end{proof} \vspace{.1in}

Qualitatively, allowing the switcher to `cheat' gives access to distributions $p \in \mC$ which
may not be in $\conv(\mG)$. Quantitatively, the conditions placed on the distributions in $\mC$
are precisely those that restrict the switcher from producing symbols that do not occur often
enough on average. For example, let $\mathcal{V} = \{1\}$ where $1 \in \mX$, and suppose that
the subsources are independent of each other. Then for every $p \in \mC$,
    \begin{equation} p(1) \geq \prod_{l=1}^m p_l(1). \end{equation}
$\prod_{l=1}^m p_l(1)$ is the probability that all $m$ subsources produce the letter $1$ at a
given time. In this case, the switcher has no option but to output the letter $1$, hence any
distribution the switcher mimics must have $p(1) \geq \prod_{l=1}^m p_l(1)$. The same logic can
be applied to all subsets $\mathcal{V}$ of $\mX$.

\section{Noisy observations of subsource realizations} \label{sec:states}

\begin{figure}
\begin{center}
\input{switcherwithstate.pstex_t}
\end{center}
\caption{A model of an AVS encompassing both cheating and non-cheating switchers. Additionally,
this model allows for noisy observations of subsource realizations by the switcher.}
\label{fig:switcherwithstate}
\end{figure}
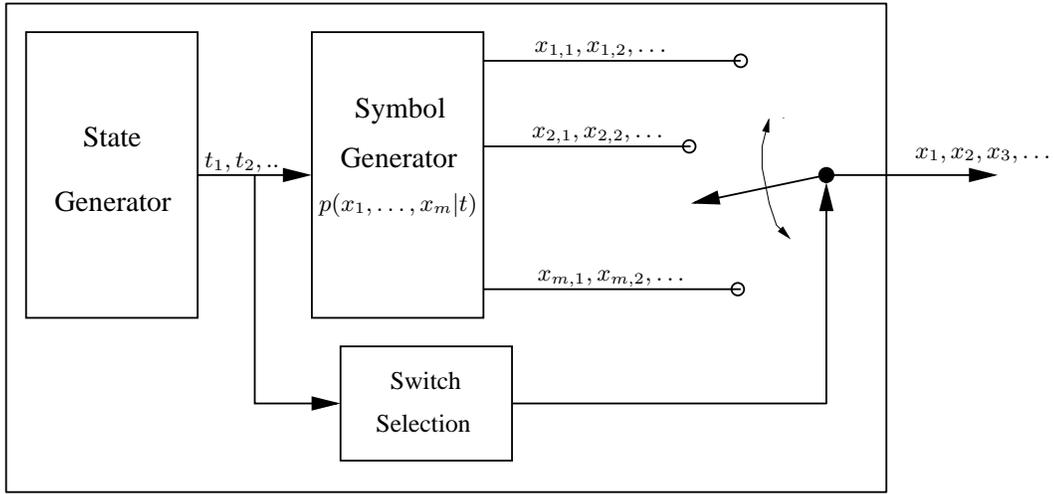

A natural extension of the AVS model is to consider the case when the adversary has noisy
access to subsource realizations through a discrete memoryless channel before pointing the
camera. Since the subsource probability distributions are already known, this model is
equivalent to one in which the switcher observes a state noiselessly. Conditioned on the state,
the $m$ subsources output symbols independent of the past according to a conditional
distribution. This model is depicted in Figure \ref{fig:switcherwithstate}.

The overall AVS is comprised now of a `state generator' and a `symbol generator' that outputs
$m$ symbols at a time. The state generator produces the state $t_k$ at time $k$ from a finite
set $\mT$. We assume the states are generated IID across time with distribution $\alpha(t)$. At
time $k$, the symbol generator outputs $(x_{1,k},\ldots, x_{m,k})$ according to
$P(x_{1,k},\ldots,x_{m,k}|t_k)$. This model allows for correlation among the subsources at a
fixed time. Let $p_l(\cdot|t), l =1,\ldots,m$, be the marginals of this joint distribution so
that conditioned on $t_k$, $x_{l,k}$ has marginal distribution $p_l(\cdot|t_k)$. For an $t \in
\mT$, let $\overline{\mG}(t) = \conv(p_1(\cdot|t), \ldots, p_m(\cdot|t))$.

The switcher can observe states either with full lookahead or $1$-step lookahead, but these two
cases will once again have the same rate-distortion function when the switcher is an adversary.
So assume that at time $k$, the switcher chooses the switch position $s_k$ with knowledge of
$\t^n, \x_1^{k-1}, \ldots, \x_m^{k-1}$. The non-cheating and cheating switcher can be recovered
as special cases of this model. If the conditional distributions $p_l(x|t)$ do not depend on
$t$, the non-cheating switcher is recovered. The cheating switcher is recovered by setting $\mT
= \mX^{m}$ and letting $p_l(x|t) = 1(x = t(l))$ where the state $t$ is an $m$ dimensional
vector consisting of the outputs of each subsource.

With this setup, we have the following extension of Theorem \ref{thm:mainresult}.

\begin{theorem} \label{thm:states}
For the AVS problem of Figure \ref{fig:switcherwithstate}, where the adversary has access to
the states either with $1$-step lookahead or full lookahead,
    \begin{equation} R(D) = \max_{p \in \mD_{states}} R(p,D), \label{eqn:states} \end{equation}
where
    \begin{equation}
    \mD_{states} = \left\{ p \in \mP(\mX) : \begin{array}{c}
         p(\cdot) = \sum_{t \in \mT} \alpha(t) f(\cdot|t)\\
        f(\cdot|t) \in \overline{\mG}(t), \forall~ t \in \mT  \end{array} \right\}. \label{eqn:stateset} \end{equation}
\end{theorem}
\begin{proof} See Appendix \ref{sec:appendixstates}. \end{proof}
\vspace{.1in}

One can see that in the case of the cheating switcher of the previous section, the set $\mD$ of
equation (\ref{eqn:alternativeset}) equates directly with $\mD_{states}$ of equation
(\ref{eqn:stateset}). In that sense, from the switcher's point of view, $\mD$ is a more natural
description of the set of distributions that can be simulated than $\mC$. Again, computing
$R(D)$ in (\ref{eqn:states}) falls into the discussion of Section \ref{sec:computingrd}.

\section{The Helpful Switcher} \label{sec:helpful}
In general, the active-source may be acting in such a way that
optimizes its own objectives.  When its objective is to output a
source sequence that is not well represented by the codebook, we
arrive at the traditional adversarial setting considered above. The
objective of the switcher, however, may vary from adversarial to
agnostic to helpful. In this section, we consider the {\em helpful}
cheating switcher. The model is as follows:
    \begin{itemize}
        \item The coder chooses a codebook that is made known to the switcher.
        \item The switcher chooses a strategy to help the coder achieve distortion $D$ on
        average with the minimum number of codewords. We consider the cases where the switcher has full
        lookahead or $1$-step lookahead.
    \end{itemize}
As opposed to the adversarial setting, a rate $R$ is now achievable at distortion $D$ if {\em
there exist} switcher strategies and codebooks for each $n$ with expected distortion at most
$D$ and the rates of the codebooks tend to $R$. The following theorem establishes $R(D)$ if the
cheating switcher has full lookahead.

\begin{theorem}\label{thm:helpfulcheating}
Let $\mX^\ast = \{ \mV\subseteq \mX: \mV \neq \emptyset, |\mV| \leq m \}$. Let $\rho: \mX^\ast
\times \mXh \to [0,d^\ast]$ be defined by
    \begin{equation} \rho(\mV, \hx) = \min_{x\in \mV} d(x,\hx). \end{equation}
Let $\mV_k = \{x_{1,k},\ldots, x_{m,k}\}$ for all $k$. Note that $\mV_i, i = 1,2, \ldots$ is a
sequence of IID random variables with distribution $\beta(\mV) = P(\{x_{1,1},\ldots,x_{m,1}\} =
\mV)$. Let $R^\ast(\beta,D)$ be the rate-distortion function for the IID source with
distribution $\beta$ at distortion $D$ with respect to the distortion measure
$\rho(\cdot,\cdot)$. For the helpful cheating switcher with full lookahead,
    \begin{equation} R(D) = R^\ast(\beta,D). \label{eqn:helpfulrd} \end{equation}
\end{theorem}
\begin{proof}
Rate-distortion problems are essentially covering problems, so we equate the rate-distortion
problem for the helpful switcher with the classical covering problem for the observed sets
$\mV_i$. If the switcher is helpful, has full lookahead, and knowledge of the codebook, the
problem of designing the codebook is equivalent to designing the switcher strategy and codebook
jointly.
    At each time $k$, the switcher observes a candidate set $\mV_k$ and must select an element
from  $\mV_k$. For any particular reconstruction codeword $\xh^n$, and a string of candidate
sets $(\mV_1, \mV_2,\ldots, \mV_n)$, the switcher can at best output a sequence $\x^n$ such
that
    \begin{equation} d_n(\x^n, \xh^n) = \frac{1}{n}\sum_{k=1}^n \rho(\mV_k, \hx_k)
    \end{equation}
Hence, for a codebook $\mB$, the helpful switcher with full lookahead can select switch
positions to output $\x^n$ such that
    \begin{equation} d_n(\x^n; \mB) = \min_{\xh^n \in \mB} \frac{1}{n}\sum_{k=1}^n \rho(\mV_k,
    \hx_k). \end{equation}
Therefore, for the helpful switcher, the problem of covering the $\mX$ space with respect to
the distortion measure $d(\cdot,\cdot)$ now becomes one of covering the $\mX^\ast$ space with
respect to the distortion measure $\rho(\cdot,\cdot)$.
\end{proof} \vspace{.1in}

{\em Remarks:}

\begin{itemize}

    \item Computing $R(D)$ in (\ref{eqn:helpfulrd}) can be done by the Blahut-Arimoto
    algorithm\cite{CoverBook}.

\item In the above proof, full lookahead was required in order
for the switcher to align the entire output word of the source with the minimum distortion
reconstruction codeword as a whole. This process cannot be done with $1$-step lookahead and so
the $R(D)$ function for a helpful switcher with $1$-step lookahead remains an open question,
but we have the following corollary of Theorems \ref{thm:mainresult} and \ref{thm:helpfulcheating}.
    \end{itemize}

\begin{corollary}\label{cor:helpful}
    For the helpful switcher with $1$-step lookahead,
        \begin{equation} R^\ast(\beta,D) \leq R(D) \leq \min_{p \in \mC} R(p,D) \end{equation}
\end{corollary}
\begin{proof} If the switcher has at least $1$-step lookahead, it immediately follows from the proof
of Theorem \ref{thm:mainresult} that $R(D) \leq \min_{p \in \mC} R(p,D)$. The question is
whether or not any lower rate is achievable. We can make the helpful switcher with $1$-step
lookahead more powerful by giving it $n$-step lookahead, which yields the lower bound
$R^\ast(\beta,D)$.
\end{proof} \vspace{.1in}

An example in Section \ref{sec:helpful_example} shows that in general, we have the
strict inequality $R^\ast(\beta,D) < \min_{p \in \mC} R(p,D)$.

One can also investigate the helpful switcher problem when the switcher has access to noisy or
partial observations as in Section \ref{sec:states}. This problem has the added flavor of
remote source coding because the switcher can be thought of as an extension of the coder and
observes data correlated with the source to be encoded. However, the switcher has the
additional capability of choosing the subsource that must be encoded. For now, this problem is
open and we can only say that $R(D) \leq \min_{p \in \mD_{states}} R(p,D)$.

\section{Examples} \label{sec:examples}
We illustrate the results with several simple examples using binary alphabets and Hamming
distortion, i.e. $\mX = \mXh = \{0,1\}$ and $d(x,\hx) = 1(x \neq \hx)$. Recall that the
rate-distortion function of an IID binary source with distribution $(p,1-p)$, $p \in
[0,\frac{1}{2}]$ is
    \begin{equation}
        R((1-p,p),D) = \left \{ \begin{array}{cc} h_b(p) - h_b(D) & D \in [0,p] \\
            0 & D > p \end{array}\right. ,
    \end{equation}
where $h_b(p)$ is the binary entropy function (in bits for this section).

\subsection{Bernoulli $1/4$ and $1/3$ sources}
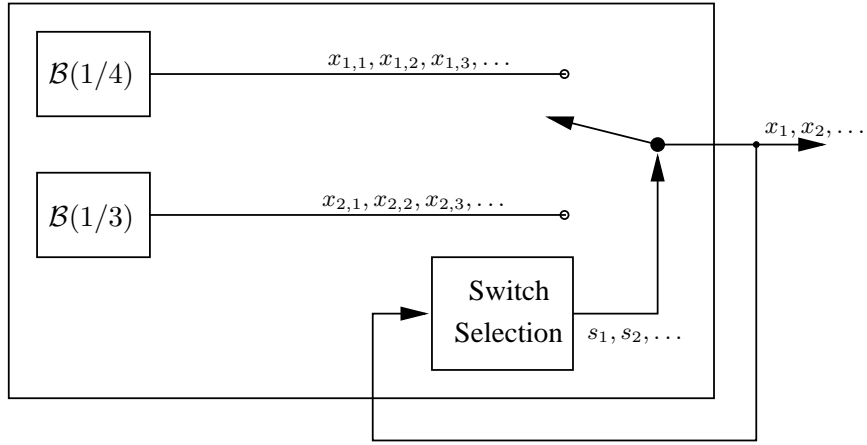
\begin{figure}[hb]
\begin{center}
    \input{example.pstex_t}
\end{center}
\caption{The adversary chooses the switch position with knowledge only of the past AVS outputs.
For Hamming distortion, the rate-distortion function is $R(D) = h_b(1/3) -h_b(D)$ for $D \in
[0,1/3]$.} \label{fig:example}
\end{figure}

\begin{figure}[hb]
\begin{center}
    \input{example_cheating.pstex_t}
\end{center}
\caption{The adversary chooses the switch position with knowledge of both subsource
realizations. For Hamming distortion, the rate-distortion function is $R(D) = 1 - h_b(D)$ for
$D \in [0,1/2]$.} \label{fig:example_cheating}
\end{figure}
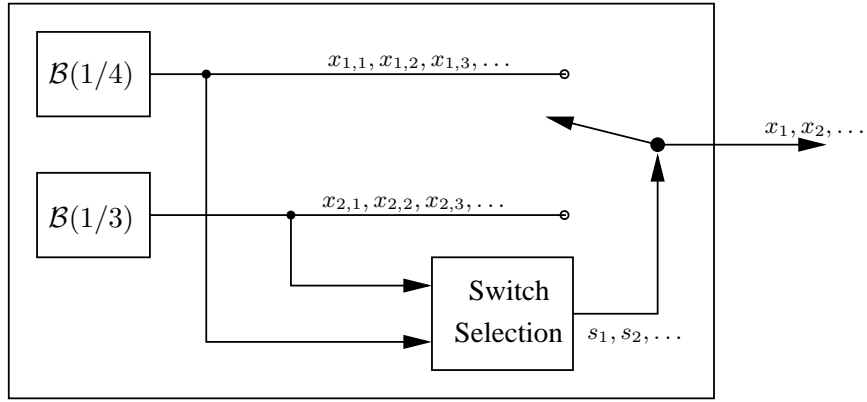

 Let $m=2$ so the switcher has access to two
IID Bernoulli subsources. Subsource $1$ outputs $1$ with probability $1/4$ and subsource $2$
outputs $1$ with probability $1/3$, so $p_1 = (3/4,1/4)$ and $p_2 = (2/3, 1/3)$. First, we
consider the switcher as an adversary. Figure \ref{fig:example} shows this example in the
traditional strictly causal setting of \cite{BergerSourceCodingGame}, where the switcher gets
only outputs of the source after the switch position has been decided. Figure
\ref{fig:example_cheating} shows the AVS in the noncausal setting, where the switcher has the
subsource realizations before choosing the switch position.

For any time $k$,
    \begin{eqnarray}
        P(x_{1,k} = x_{2,k} = 0) & = & \frac{3}{4}\cdot\frac{2}{3} = \frac{1}{2} \\
        P(x_{1,k} = x_{2,k} = 1) & = & \frac{1}{4}\cdot\frac{1}{3} = \frac{1}{12}\\
        P(\{x_{1,k}, x_{2,k}\} = \{0,1\}) & = & 1 - \frac{1}{2} - \frac{1}{12} = \frac{5}{12}.
    \end{eqnarray}

\begin{figure}
\begin{center}
\input{example_line_simplex.pstex_t}
\end{center}
\caption{The binary distributions the switcher can mimic. $\conv(\mG)$ is the set of
distributions the switcher can mimic with causal access to subsource realizations, and $\mC$ is
the set attainable with noncausal access.} \label{fig:example_simplex}
\end{figure}
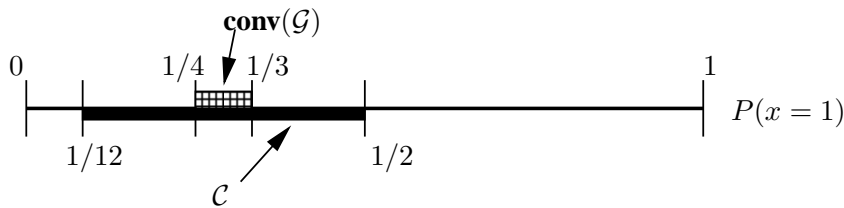

If the switcher is allowed $1$-step lookahead and has the option of choosing either $0$ or $1$,
suppose the switcher chooses $1$ with probability $f_1$. The coder then sees an IID binary
source with a probability of a $1$ occurring being equal to:
    \begin{equation}
        p(1) = \frac{1}{12} + \frac{5}{12} f_1.
    \end{equation}
By using $f_1$ as a parameter, the switcher can produce $1$'s with any probability between
$1/12$ and $1/2$. The attainable distributions are shown in Figure \ref{fig:example_simplex}.
The switcher with lookahead can simulate a significantly larger set of distributions than the
causal switcher, which is restricted to outputting $1$'s with probability in $[1/4,1/3]$. Thus,
for the strictly causal switcher, $R(D) = h_b(1/3) - h_b(D)$ for $D \in [0,1/3]$ and for the
switcher with $1$-step or full lookahead, $R(D) = 1 - h_b(D)$ for $D \in [0,1/2]$.

\begin{figure}
\begin{center}
    \input{example_mod2.pstex_t}
\end{center}
\caption{The adversary observes the mod-2 sum of the two subsources, a Bernoulli $1/3$ subsource and a
Bernoulli $1/4$ subsource. For Hamming distortion, the rate-distortion function is $R(D) =
h_b(1/3) - h_b(D)$ for $ D \in [0,1/3]$.} \label{fig:example_mod2}
\end{figure}
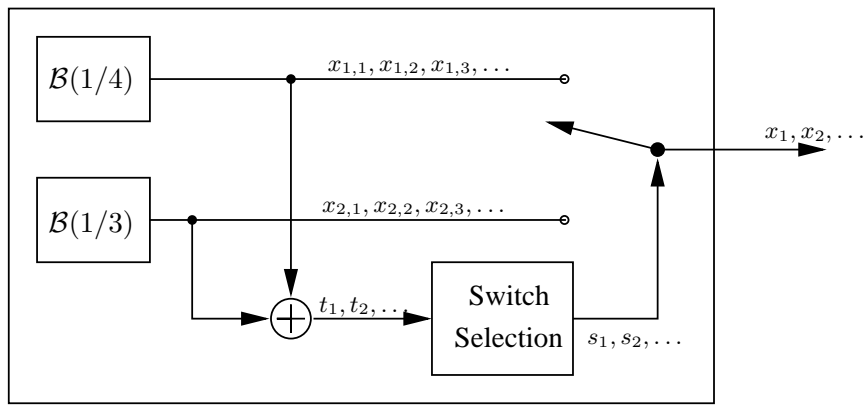

We now look at several variations of this example to illustrate the utility of noisy or partial
observations of the subsources for the switcher. In the first variation, shown in Figure
\ref{fig:example_mod2}, the switcher observes the mod-$2$ sum of the two subsources. Theorem
\ref{thm:states} then implies that $R(D) = h_b(1/3) - h_b(D)$ for $D \in [0,1/3]$. Hence, the
mod-$2$ sum of these two subsources is useless to the switcher in deciding the switch position.
This is intuitively clear from the symmetry of the mod-$2$ sum. If $t = 0$, either both
subsources are $0$ or both subsources are $1$, so the switch position doesn't matter in this
state. If $t = 1$, one of the subsources has output $1$ and the other has output $0$, but
because of the symmetry of the mod-$2$ function, the switcher's prior as to which subsource
output the $1$ does not change and it remains that subsource $2$ was more likely to have output
the $1$.

\begin{figure}[hb]
\begin{center}
    \input{example_x2.pstex_t}
\end{center}
\caption{The adversary observes the second subsource perfectly, but does not observe the first
subsource. For Hamming distortion, the rate-distortion function is $R(D) = 1 - h_b(D)$ for $D \in
[0,1/2]$.} \label{fig:example_x2}
\end{figure}
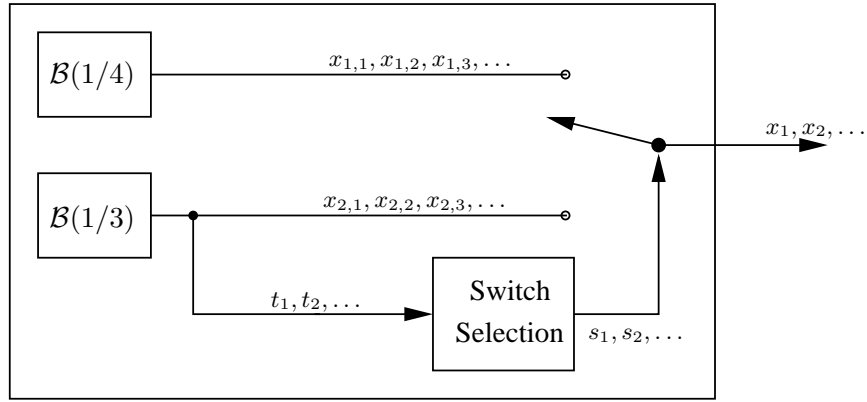

In the second variation, shown in Figure \ref{fig:example_x2}, the switcher observes the second
subsource directly but not the first, so $t_k = x_{2,k}$ for all $k$. Using Theorem
\ref{thm:states} again, it can be deduced that in this case $R(D) = 1 - h_b(D)$ for $D \in
[0,1/2]$. This is also true if $t_k = x_{1,k}$ for all $k$, so observing just one of the
subsources noncausally is as beneficial to the switcher as observing both subsources
noncausally. This is clear in this example because the switcher is attempting to output as many
$1$'s as possible. If $t = 1$, the switcher will set the switch position to $2$ and if $t = 0$,
the switcher will set the switch position to $1$ as there is still a chance that the first
subsource outputs a $1$.

\begin{figure}[hb]
\begin{center}
    \includegraphics[width = 5in]{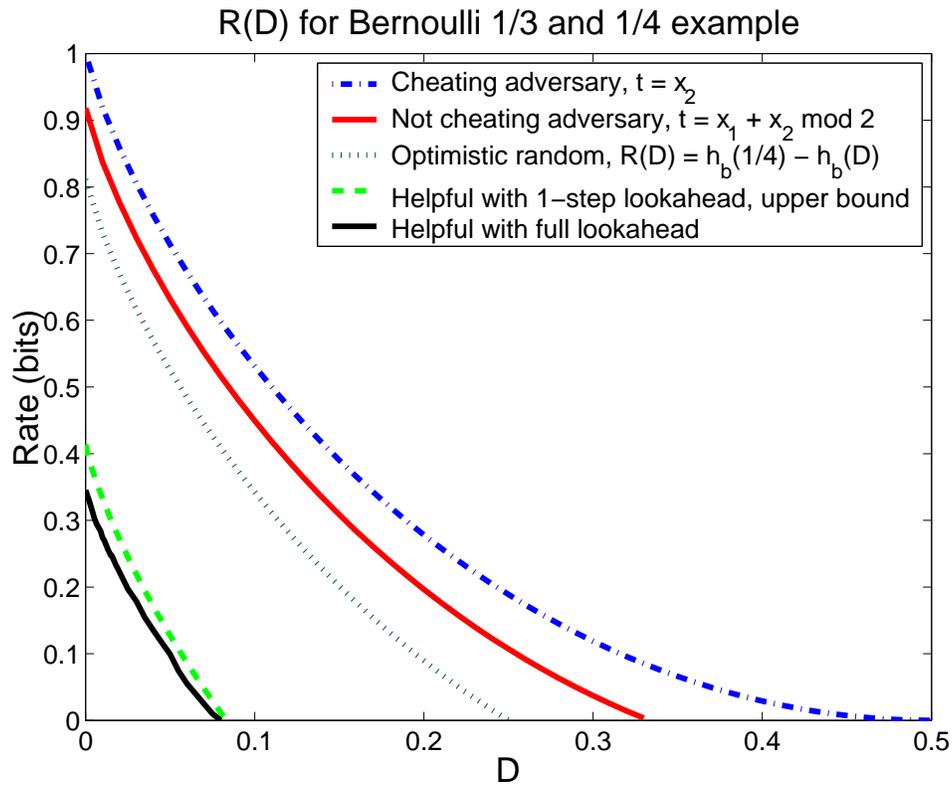}
\end{center}
\caption{$R(D)$ for the cheating switcher and the non-cheating switcher. Also, the
rate-distortion function for the examples of Figures \ref{fig:example_mod2} and
\ref{fig:example_x2}.} \label{fig:cheatingvsnot}
\end{figure}

For this example, the helpful cheater with $1$-step lookahead has a rate-distortion function
that is upper bounded by $h_b(1/12) - h_b(D)$ for $D \in [0,1/12]$. The rate-distortion
function for the helpful cheater with full lookahead can be computed from Theorem
\ref{thm:helpfulcheating}. In Figure \ref{fig:cheatingvsnot}, the rate-distortion function is
plotted for the situations discussed so far. In an active sensing situation, we see that there
can be a large gap between the required rates for adversarially modelled active sensors and
sensors which have been jointly optimized with the coding system.

\clearpage

\begin{figure}
\begin{center}
    \input{example_bsc.pstex_t}
\end{center}
\caption{The adversary observes the second subsource transmitted over a binary symmetric channel
with crossover probability $\delta$. For Hamming distortion, the rate-distortion function is
$R(D) = h_b(1/3) - h_b(D)$ for $D \in [0,1/3]$ if $\delta \in [2/5,1/2]$. If $\delta \in
[0,2/5)$, $R(D) = h_b(1/2 - 5\delta/12) - h_b(D)$ for $D \in [0, 1/2 - 5\delta/12]$.}
\label{fig:example_bsc}
\end{figure}
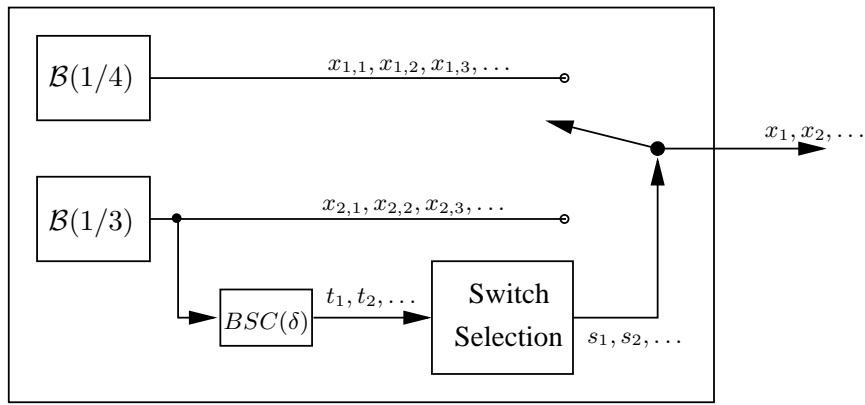

\begin{figure}
\begin{center}
    \includegraphics[width = 3in]{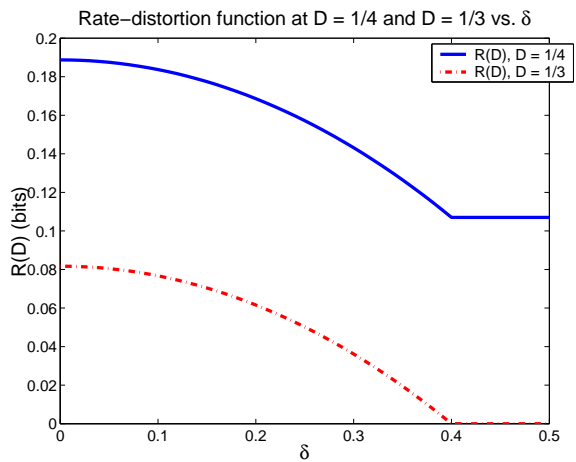}
\end{center}
\caption{$R(D)$ as a function of the noisy observation crossover probability $\delta$ for two different distortions for the example of
Figure \ref{fig:example_bsc}.} \label{fig:ratevsdelta}
\end{figure}

Finally, in Figure \ref{fig:example_bsc}, an adversarial switcher observes the second subsource
through a binary symmetric channel with crossover probability $\delta \in [0,1/2]$. Applying
Theorem \ref{thm:states} again, it can be shown that if $\delta \in [0,2/5]$,
    \begin{equation} R(D) = h_b\l( \frac{1}{2} - \frac{5}{12}\delta \r) - h_b(D),~D\in \l[ 0,
    \frac{1}{2} - \frac{5}{12}\delta \r] \end{equation}
and if $\delta \in [2/5,1/2]$,
    \begin{equation} R(D) = h_b\l(\frac{1}{3}\r) - h_b(D), ~ D \in \l[0, \frac{1}{3}\r].
    \end{equation}
Here, increasing $\delta$ decreases the switcher's knowledge of the subsource realizations.
Somewhat surprisingly, the utility of the observation is exhausted at $\delta = 2/5$, even
before the state and observation are completely independent at $\delta = 1/2$. This can be
explained through the switcher's {\em a posteriori} belief that second subsource output was a
$1$ given the state. If the switcher observes $t=1$ and $\delta \leq 1/2$, $p(x_{2,k} = 1|t_k
=1 ) \geq 1/3 > 1/4$ so the switch position will be set to $2$. When the switcher observes $t =
0$, if $\delta \leq 2/5$, $p(x_{2,k} = 1|t_k = 0) \leq 1/4$, so the switch will be set to
position $1$. However, if $\delta > 2/5$, $p(x_{2,k} = 1| t_k = 0) > 1/4$, so the switch
position will be set to $2$ even if $t = 0$ because the switcher's {\em a posteriori} belief is
that the second subsource is {\em still} more likely to have output a $1$ than the first
subsource. Figure \ref{fig:ratevsdelta} shows $R(D)$ for this example as a function of $\delta$
for two values of $D$.

\subsection{Two Bernoulli $1/2$ subsources} \label{sec:helpful_example}
\begin{figure}
\begin{center}
    \includegraphics[width = 3in]{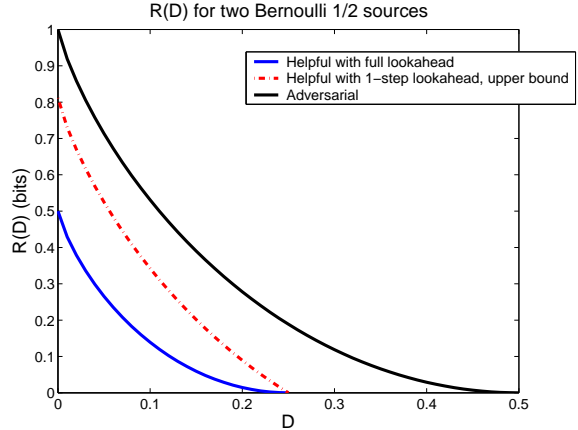}
\end{center}
\caption{The $R(D)$ function for a helpful switcher with full lookahead. For $1$-step
lookahead, the upper bound is shown.} \label{fig:example_helpful}
\end{figure}

    Suppose $m=2$, and both subsources are Bernoulli $1/2$ IID processes. For this example,
the rate-distortion function is $R(D) = 1 - h_b(D)$ for $D \in [0,1/2]$ whether the adversarial
switcher is strictly causal, causal or noncausal. When the helpful switcher has $1$-step
lookahead, $R(D) \leq R_U(D) = h_b(1/4) - h_b(D)$ for $D \in [0,1/4]$. One can also think of
this upper bound as being the rate-distortion function for the helpful switcher with $1$-step
lookahead that is restricted to using memoryless, time-invariant rules. Using Theorem $9.4.1$
of \cite{GallagerBook}, one can show that when the switcher has full lookahead,
    \begin{equation} R(D) = R^\ast(\beta,D) = \frac{1}{2}\l [ 1 - h_b(2D)\r], ~ D \in [0,1/4].
    \end{equation}
The plot of these functions in Figure \ref{fig:example_helpful} shows that the rate-distortion
function can be significantly reduced if the helpful switcher is allowed to observe the entire
block of subsource realizations.
    It is also interesting to note {\em how} the switcher with full lookahead helps the coder
achieve a rate of $R^\ast(\beta,D)$. In this example $\mX^\ast = \{ \{0\}, \{1\}, \{0,1\}\}$,
$\rho(\{0\},\hx) =1(0 \neq \hx)$, $\rho(\{1\},\hx) = 1(1 \neq \hx)$, $\rho(\{0,1\},\hx) = 0$
and $\beta = (1/4, 1/4, 1/2)$. The $R^\ast(\beta,D)$ achieving distribution on $\mXh$ is $(1/2,
1/2)$, but $R^\ast(\beta,D) < 1 - h_b(D)$. The coder is attempting to cover strings with types
near $(1/2,1/2)$ but with far fewer codewords than are needed to do so. This problem is
circumvented through the aid provided by the switcher in pushing the output of the source
inside the Hamming $D$-ball of a codeword. This is in contrast to the strategy that achieves
$R_U(D)$, where the switcher makes the output an IID sequence with as few $1$'s as possible and
the coder is expected to cover {\em all} strings with types near $(3/4,1/4)$.

\section{Computing $R(D)$ for an AVS} \label{sec:computingrd}

The $R(D)$ function for an AVS with either causal or noncausal access
to the subsource realizations is of the form
        \begin{equation} R(D) = \max_{p \in \mQ} R(p,D), \label{eqn:computation} \end{equation}
where $\mQ$ is a set of distributions in $\mP(\mX)$. In (\ref{eqn:bergerscgamerd}),
(\ref{eqn:defnC}), and (\ref{eqn:stateset}) $\mQ$ is defined by a finite number of linear
inequalities and hence is a polytope. The number of constraints in the definition of $\mQ$ is
exponential in $|\mX|$ or $|\mT|$ when the adversary has something other than strictly causal
knowledge. Unfortunately, the problem of finding $R(D)$ is not a convex program because
$R(p,D)$ is not a concave $\cap$ function of $p$ in general. In fact, $R(p,D)$ may not even be
quasi-concave and may have multiple local maxima with values different from the global maximum
as shown by Ahlswede \cite{AhlswedeExtremalProperties}.

Since standard convex optimization tools are unavailable for this problem, we consider the
question of how to approximate $R(D)$ to within some (provable) precision. That is, for any $\eps
> 0$, we will consider how to provide an approximation $R_a(D)$ such that $|R_a(D) - R(D)| \leq \eps$. Note that for fixed $p$, $R(p,D)$ can be computed efficiently by the
Blahut-Arimoto algorithm to any given precision, say much less than $\eps$. Therefore, we
assume that $R(p,D)$ can be computed for a fixed $p$ and $D$. We also assume $D \geq
D_{\min}(\mQ)$ since otherwise $R(D) = \infty$. Checking this condition is a linear program
since $\mQ$ is a polytope and $D_{\min}(p)$ is linear in $p$.

We will take a `brute-force' approach to computing $R(D)$. That is, we
wish to compute $R(p,D)$ for (finitely) many $p$ and then maximize
over the computed values to yield $R_a(D)$. Since $R(p,D)$ is
uniformly continuous in $(p,D)$ and hence in $p$, it is possible to do
this and have $|R_a(D) - R(D)| \leq \eps$ provided enough
distributions $p$ are `sampled'.  Undoubtedly, there are other
algorithms to compute $R(D)$ that likely have better problem-size
dependence. In this section, we are only interested in showing that
$R(D)$ can provably be computed to within any required precision with
a finite number of computations.

\subsection{Uniform continuity of $R(p,D)$}
The main tool used to show that the rate-distortion function can be
approximated is an explicit bound on the uniform continuity of
$R(p,D)$ in terms of $\|p-q\|_1 = \sum_{x\in \mX} |p(x) - q(x)|$ for
distortion measures that allow for $0$-distortion to be achieved
regardless of the source. In \cite{CoverBook}, a bound on the
continuity of the entropy of a distribution is developed in terms of
$\|p-q\|_1$.

\begin{lemma}[$\mL_1$ bound on continuity of entropy \cite{CoverBook}] \label{lem:entropybound} Let $p$ and
$q$ be two probability distributions on $\mX$ such that $\|p-q\|_1 \leq 1/2$, then
    \begin{equation} |H(p) - H(q)| \leq \|p-q\|_1 \ln \frac{|\mX|}{\|p-q\|_1}. \end{equation}
\end{lemma}
\vspace{.1in}

In the following lemma, a similar uniform continuity is stated for $R(p,D)$. The proof makes
use of Lemma \ref{lem:entropybound}.

\begin{lemma}[Uniform continuity of $R(p,D)$]\label{lem:ratedistortionbound}
Let $d: \mX \times \mXh \rightarrow [0,d^\ast]$ be a distortion function. $\widetilde{d}$ is
the minimum nonzero distortion from (\ref{eqn:mindistortion}). Also, assume that for each $x \in
\mX$, there is an $\hat{x}_0(x) \in \mXh$ such that $d(x, \hat{x}_0(x)) = 0$. Then, for $p, q
\in \mP(\mX)$ with $\|p - q\|_1 \leq \frac{\widetilde{d}}{4d^\ast}$, for any $D \geq 0$,
    \begin{equation} |R(p,D) - R(q,D)| \leq \frac{7d^\ast}{\widetilde{d}} \| p -
    q\|_1 \ln \frac{|\mX||\mXh|}{\|p - q\|_1}. \label{eqn:result} \end{equation}
\end{lemma}
\begin{proof} See Appendix \ref{sec:appendix2}.\end{proof}
\vspace{.1in}

The restriction that $d(x,\cdot)$ has at least one zero for every $x$ can be relaxed if we are
careful about recognizing when $R(p,D)$ is infinite. For an arbitrary distortion measure $d:
\mX \times \mXh \to [0, d^\ast]$, define
    \begin{equation}
        d_0(x,\hx) = d(x,\hx) - \min_{\wt{x} \in \mXh} d(x,\wt{x}).
    \end{equation}
Now let $d_0^\ast = \max_{x, \hx} d_0(x,\hx)$ and $\wt{d}_0 = \min_{(x,\hx): d_0(x,\hx)>0}
d_0(x,\hx)$. We have defined $d_0(x,\hx)$ so that Lemma \ref{lem:ratedistortionbound} applies,
so we can prove the following lemma.

\begin{lemma}\label{lem:ratedistortionbound2} Let $p,q\in \mP(\mX)$ and let $D \geq \max(D_{\min} (p), D_{\min} (q))$.
If $\|p-q\|_1 \leq \wt{d}_0/4d^\ast$,
    \begin{equation} |R(p,D) - R(q,D)| \leq \frac{11 d^\ast}{\wt{d}_0} \|p-q\|_1 \ln
    \frac{|\mX||\mXh|}{\|p-q\|_1}.
    \end{equation}
\end{lemma}
\begin{proof} See Appendix \ref{sec:appendix3}. \end{proof} \vspace{.1in}

As $\|p-q\|_1$ goes to $0$, $-\ln \|p-q\|_1$ goes to infinity slowly and it can be shown that
for any $\delta \in (0,1)$ and $\gamma \in [0,1/2]$,
    \begin{equation} \gamma \ln \frac{|\mX||\mXh|}{\gamma} \leq \frac{(|\mX||\mXh|)^\delta}{e\delta} \gamma^{1-\delta}.
    \label{eqn:functionfact}
    \end{equation}

In the sequel, we let $f(\gamma) = \gamma \ln \frac{|\mX||\mXh|}{\gamma}$ for $\gamma \in [0,
1/2]$ with $f(0) = 0$ by continuity. It can be checked that $f$ is strictly monotonically
increasing and continuous on $[0,1/2]$ and hence has an inverse function $g: f([0,1/2]) \to [0,
1/2]$, i.e. $g(f(\gamma)) = \gamma$ for all $\gamma \in [0,1/2]$. Note that $g$ is not
expressible in a simple `closed-form', but can be computed numerically.

\subsection{A bound on the number of distributions to sample}
Returning to the problem of computing $R(D)$ in equation (\ref{eqn:computation}), consider the
following simple algorithm. Without loss of generality, assume $\mX = \{1,2,\ldots, |\mX|\}$. Let $\gamma \in (0,1)$
and let $\gamma \Z^{|\mX|-1}$ be the $|\mX|-1$ dimensional integer lattice scaled by $\gamma$.
Let $\wt{\mathcal{O}} = [0,1]^{|\mX|-1} \bigcap \gamma \Z^{|\mX|-1}$. Now, define
    \begin{equation} \mathcal{O} = \l \{ q \in \mP(\mX): \begin{array}{c} \exists ~\wt{q} \in
    \wt{\mO}, \\ q(i) = \wt{q}(i), i = 1,\ldots, |\mX|-1, \\ q(|\mX|) = 1 - \sum_{i
    =1}^{|\mX|-1} \wt{q}(i) \geq 0 \end{array} \r \}. \end{equation}

In words, sample the $|\mX|-1$ dimensional unit cube, $[0,1]^{|\mX|-1}$, uniformly with points
from a scaled integer lattice. Embed these points in $\R^{|\mX|}$ by assigning the last value
of the new vector to be $1$ minus the sum of the values in the original point. If this last
value is non-negative, the new point is a distribution in $\mP(\mX)$. The algorithm to compute
$R_a(D)$ is then one where we compute $R(p,D)$ for distributions $q \in \mO$ that are also in
or close enough to $\mQ$.
    \begin{enumerate}
        \item Fix a $q \in \mO$. If $ \min_{p \in \mQ} \|p - q\|_1 \leq 2|\mX|\gamma$, compute $R(q,D)$,
        otherwise do not compute $R(q,D)$. Repeat for all $q \in \mO$.

        \item Let $R_a(D)$ be the maximum of the computed values of $R(q,D)$, i.e.
            \begin{equation} R_a(D) = \max\l\{R(q,D): q \in \mO, \min_{p \in \mQ} \|p-q\|_1 \leq
            2|\mX|\gamma \r \}. \end{equation}
    \end{enumerate}
Checking the condition $\min_{p\in \mQ} \|p - q\|_1 \leq \gamma 2|\mX|$ is essentially a linear program, so
it can be efficiently solved. By setting $\gamma$ according to the accuracy $\eps > 0$ we want,
we get the following result.

\begin{theorem} The preceding algorithm computes an approximation $R_a(D)$ such that $|R_a(D) -
R(D)| \leq \eps$ if
    \begin{equation} \gamma \leq \frac{1}{2|\mX|} g\l(\frac{\eps \wt{d_0}}{11d^\ast} \r).
    \end{equation}
The number of distributions for which $R(q,D)$ is computed to determine $R(D)$ to within
accuracy $\eps$ is at most\footnote{This is clearly not the best bound as many of the points in
the unit cube on do not yield distributions on $\mP(\mX)$. The factor by which we are
overbounding is roughly $|\mX|!$, but this factor does not affect the dependence on $\eps$.}
    \begin{equation} N(\eps) \leq \l ( \frac{2|\mX|}{g\l(\frac{\eps \wt{d}_0}{11 d^\ast} \r)} +
    2 \r)^{|\mX| - 1}. \end{equation}

\end{theorem}
\begin{proof} The bound on $N(\eps)$ is clear because the number of points in $\wt{\mO}$ is at
most $(\lceil 1/\gamma \rceil + 1)^{|\mX|-1}$ and every distribution in $\mO$ is associated
with one in $\wt{\mO}$, so $|\mO| \leq |\wt{\mO}|$.

Now, we prove $|R_a(D) - R(D)| \leq \eps$. For this discussion, we let $\gamma
=\frac{1}{2|\mX|} g\l(\frac{\eps \wt{d_0}}{11d^\ast}\r)$. First, for all $p \in \mQ$, there is
a $q \in \mO$ with $\|p - q\|_1 \leq g\l(\frac{\eps \wt{d}_0}{11 d^\ast} \r) = 2|\mX|\gamma$.
To see this, let $\wt{q}(i) = \lfloor \frac{p(i)}{\gamma}\rfloor \gamma$ for $i = 1, \ldots,
|\mX|-1$. Then $\wt{q} \in \wt{\mO}$, and we let $q(i) = \wt{q}(i)$ for $i = 1,\ldots,
|\mX|-1$. Note that
    \begin{equation} q(|\mX|) = 1 - \sum_{i =1 }^{|\mX|-1} q(i) = 1 - \sum_{i=1}^{|\mX|-1}
    \l\lfloor \frac{p(i)}{\gamma} \r\rfloor \gamma \geq 1 - \sum_{i=1}^{|\mX|-1} p(i) = p(|\mX|)\geq 0. \end{equation}
Therefore $q \in \mO$ and furthermore,
    \begin{eqnarray} \|p - q\|_1 & \leq & \l(1 - \sum_{i=1}^{|\mX|-1} (p(i) - \gamma) - p(|\mX|)\r) + \sum_{i=1}^{|\mX|-1} \l( p(i) -  \l\lfloor \frac{p(i)}{\gamma} \r\rfloor
    \gamma \r )\\
        & \leq & 2(|\mX|-1)\gamma\\
        & \leq & 2|\mX| \gamma \\
        & \leq & g\l(\frac{\eps \wt{d}_0}{11 d^\ast} \r).
    \end{eqnarray}
By Lemma \ref{lem:ratedistortionbound2}, $R(q,D) \geq R(p,D) - \eps$. This distribution $q$ (or
possibly one closer to $p$) will always be included in the maximization yielding $R_a(D)$, so
we have $R_a(D) \geq \max_{p \in \mQ} R(p,D) - \eps = R(D) - \eps$.

Conversely, for a $q \in \mO$, if $\min_{p \in \mQ} \|p-q\|_1 \leq 2|\mX|\gamma$, Lemma
\ref{lem:ratedistortionbound2} again gives
    \begin{equation} R(q,D) \leq \max_{p \in \mQ} R(p,D) + \eps = R(D) + \eps \end{equation}
Therefore, $|R_a(D) - R(D)| \leq \eps$.
\end{proof}
\vspace{.1in}

\subsection{Estimation of the rate-distortion function of an unknown IID source}
\label{sec:estimation} An explicit bound on the continuity of the rate-distortion function has
other applications. Recently, Harrison and Kontoyiannis \cite{HarrisonEstimationRateDistortion}
have studied the problem of estimating the rate-distortion function of the marginal
distribution of an unknown source. Let $p_{\x^n}$ be the (marginal) empirical distribution of a
vector $\x^n \in \mX^n$. They show that the `plug-in' estimator $R(p_{\x^n},D)$, the
rate-distortion function of the empirical marginal distribution of a sequence, is a consistent
estimator for a large class of sources beyond just IID sources with known alphabets. However,
if the source is known to be IID with alphabet size $|\mX|$, estimates of the convergence rate
(in probability) of the estimator can be provided using the uniform continuity of the
rate-distortion function.

Suppose the true source is IID with distribution $p \in \mP(\mX)$ and fix a probability $\tau
\in (0,1)$ and an $\eps \in (0,\ln|\mX|)$. We wish to answer the question: How many samples $n$
need to be taken so that $|R(p_{\x^n},D) - R(p,D)| \leq \eps$ with probability at least
$1-\tau$? The following lemma gives a sufficient number of samples $n$.

\begin{theorem}\label{lem:estimation}
Let $d: \mX \times \mXh \to [0,d^\ast]$ be a distortion measure for which Lemma
\ref{lem:ratedistortionbound} holds. For any $p \in \mP(\mX)$, $\tau \in (0,1)$, and $\eps \in
(0, \ln |\mX|)$, then
    \begin{equation} P(|R(p_{\x^n},D) - R(p,D)| \geq \eps) \leq \tau \end{equation}
if
    \begin{equation} n > \frac{2}{g\l( \frac{\eps \wt{d}}{7 d^\ast} \r)^2} \l ( \ln \frac{1}{\tau} + |\mX| \ln
    2\r). \label{eqn:sufficientsamples}
    \end{equation}
\end{theorem}
\begin{proof}
From Lemma \ref{lem:ratedistortionbound}, we have
    \begin{eqnarray}
        P(|R(p_{\x^n},D) - R(p,D)| \geq \eps) & \leq & P\l(\|p_{\x^n} - p\|_1 \geq g\l (\frac{\eps \wt{d}}{ 7d^\ast}
        \r)\r)\\
        & \leq & 2^{|\mX|} \exp\l(-\frac{n}{2}g\l(\frac{\eps \wt{d}}{7 d^\ast}
        \r)^2 \r)
    \end{eqnarray}
The last line follows from Theorem 2.1 of \cite{L1Deviation}. This bound is similar to, but a
slight improvement over, the method-of-types bound of Sanov's Theorem. Rather than an
$(n+1)^{|\mX|}$ term, we just have a $2^{|\mX|}$ term multiplying the exponential. Taking $\ln$
of both sides gives the desired result.
\end{proof}

We emphasize that this number $n$ is a sufficient number of samples regardless of what the true
distribution $p \in \mP(\mX)$ is. The bound of (\ref{eqn:sufficientsamples}) depends only on
the distortion measure $d$, alphabet sizes $|\mX|$ and $|\mXh|$, desired accuracy $\eps$ and
`estimation error' probability $\tau$.

\section{Concluding Remarks} \label{sec:conclusion}
   As mentioned in the introduction, the active-source problem is truly interesting when the sources have memory.
Dobrushin \cite{DobrushinMemory} has analyzed the case of the non-anticipatory AVS composed of
independent sources with memory with different distributions when the switcher is passive and
blindly chooses the switch position. In the case of sources with memory, additional
knowledge will no doubt increase the adversary's power to increase the rate-distortion function. If we let $R^{(k)}(D)$ be the rate-distortion function for an AVS
composed of sources with memory and an adversary with $k$ step lookahead, one could imagine
that in general,
    \begin{equation}
        R^{(0)}(D) < R^{(1)}(D) < R^{(2)}(D) < \cdots < R^{(\infty)}(D).
    \end{equation}

Another interesting problem, at least mathematically, is the
arbitrarily varying channel formulation analogous to the problems of
Sections \ref{sec:result} and \ref{sec:states}.  Similar techniques to
those developed here might prove useful in considering a cheating
`jammer' for an arbitrarily varying channel. While the problem is well
defined, it seems unphysical in the usual context of jamming or
channel noise. The idea may make more sense in the context of
watermarking, where the adversary can try many different attacks on
different letters of the input before deciding to choose one for each.

For the original motivation of compressing active-vision sources, the results here suggest that
treating it as an adversarial black box might be overly conservative. There is a large gap
between the adversarial and helpful rate-distortion functions. This suggests that an
interesting question to study is one of mismatched objectives where the switcher is trying to
be helpful for some particular distortion metric but the source is actually being encoded with
a different metric in mind. Finally, if the active-sensor and coding system are part of a
tightly delay-constrained control loop, we would want to study these issues from the causal
source code perspective of \cite{NeuhoffGilbert}. It seems likely that the adversarial results
of Theorems \ref{thm:mainresult} and \ref{thm:states} would follow straightforwardly with the
same sets of distributions $\mC$ and $\mD$, with the IID rate-distortion function for noncausal
source codes replaced by the the IID rate-distortion functions for causal source codes.

\appendices

\section{Proof of Theorem \ref{thm:mainresult}}
\label{sec:appendix1}
\subsection{Achievability for the coder} The main tool of the proof is:

\begin{lemma}[Type Covering] Let $S_D(\xh^n) \triangleq \{ \x^n \in \mX^n : d_n(\x^n,\xh^n) \leq D\}$ be the set of $\mX^n$
strings that are within distortion $D$ of a $\mXh^n$ string $\xh^n$. Fix a $p \in \mP_n(\mX)$
and an $\epsilon > 0$. Then for all $n$ large enough, there exist codebooks $\mB = \{\xh^n(1),
\xh^n(2), \ldots, \xh^n(M)\}$ where $M < \exp(n(R(p,D) + \epsilon))$ and
    \begin{equation} T_p^n \subseteq \bigcup_{\xh^n \in \mB} S_D(\xh^n), \end{equation}
where $T_p^n$ is the set of $\mX^n$ strings with type $p$.
\end{lemma}\vspace{.1in}

\begin{proof} See \cite{BergerSourceCodingGame}, Lemma 1. \end{proof}

We now show how the coder can get arbitrarily close to $\widetilde{R}(D)$ for large enough $n$.
For $\delta > 0$, define $\mC_\delta$ as
    \begin{equation}
        \mC_\delta \triangleq \left\{ \begin{array}{ccc}
        & & \sum_{x\in \mathcal{V}} p(x) \geq  P(x_l \in \mV, 1 \leq l \leq m) - \delta \\
        p \in \mP(\mX) & : & \forall~ \mathcal{V} \textrm{ such that } \\
        & & \mathcal{V} \subseteq \mX
        \end{array} \right\}.
    \end{equation}

\begin{lemma}[Converse for switcher]
Let $\epsilon > 0$. For all $n$ sufficiently large \begin{equation}\frac{1}{n}\ln{M(n,D)} \leq
\widetilde{R}(D) + \epsilon. \end{equation}
\end{lemma} \vspace{.1in}

\begin{proof}
We know $R(p,D)$ is a continuous function of $p$ (\cite{CsiszarBook}). It follows then that
because $\mC_\delta$ is monotonically decreasing (as a set) with $\delta$ that for all
$\epsilon > 0$, there is a $\delta > 0$ so that
    \begin{equation}
        \max_{p \in \mC_\delta} R(p,D) \leq \max_{p \in \mC} R(p,D) + \epsilon/2.
    \end{equation}

We will have the coder use a codebook such that all $\mX^n$ strings with types in $\mC_\delta$
are covered within distortion $D$. The coder can do this for large $n$ with at most $M$
codewords in the codebook $\mB$, where
    \begin{eqnarray}
        M & < & (n+1)^{|\mX |}\exp(n\max_{p \in \mC_\delta} R(p,D)) \\
        & \leq & \exp(n(\max_{p\in \mC} R(p,D) + \epsilon)).
    \end{eqnarray}

Explicitly, this is done by taking a union of the codebooks provided by the type-covering lemma
and noting that the number of types in $\mP_n(\mX)$ is less than $(n+1)^{|\mX|}$. Next, we will
show that the probability of the switcher being able to produce a string with a type not in
$\mC_\delta$ goes to $0$ exponentially with $n$.

Consider a type $p \in \mP_n(\mX) \cap (\mP(\mX) - \mC_\delta)$. By definition, there is some
$\mathcal{V} \subseteq \mX$ such that $\sum_{x\in \mathcal{V}} p(x) < P(x_l \in \mV, 1 \leq l
\leq m) - \delta$. Let $\zeta_k(\mathcal{V})$ be the indicator function
    \begin{equation} \zeta_k(\mathcal{V}) = \prod_{l=1}^m \1(x_{l,k} \in \mathcal{V}). \end{equation}
$\zeta_k$ indicates the event that the switcher cannot output a symbol outside of $\mathcal{V}$
at time $k$. Then $\zeta_k(\mathcal{V})$ is a Bernoulli random variable with a probability of
being $1$ equal to $\kappa(\mV) \triangleq P(x_l \in \mV, 1 \leq l \leq m)$. Since the
subsources are IID over time, $\zeta_k(\mathcal{V})$ is a sequence of IID binary random
variables with distribution $q' \triangleq (1-\kappa(\mV), \kappa(\mV))$.

Now for the type $p \in \mP_n(\mX) \cap (\mP(\mX) - \mC_\delta)$, we have that for all strings
$\x^n$ in the type class $T_p$, $\frac{1}{n} \sum_{i=1}^n \1(x_i \in \mathcal{V}) < \kappa(\mV)
- \delta$. Let $p'$ be the binary distribution $(1 - \kappa(\mV) + \delta, \kappa(\mV) -
\delta)$. Therefore $||p' - q'||_1 = 2\delta$, and hence we can bound the binary divergence
$D(p'||q') \geq 2\delta^2$ by Pinsker's inequality. Using standard types properties
\cite{CoverBook} gives
    \begin{eqnarray}
        P\bigg ( \frac{1}{n} \sum_{k=1}^n \zeta_k(\mathcal{V}) < \kappa(\mV) - \delta \bigg) & \leq & (n+1) \exp(-nD(p'||q')) \\
        & \leq & (n+1) \exp(-2n\delta^2).
    \end{eqnarray}

This bound holds for all $\mV \subset \mX, \mV \neq \emptyset$, so we sum over types not in
$\mC_\delta$ to get
    \begin{eqnarray}
        P(p_{\x^n} \notin \mC_\delta)& \leq & \sum_{p \in \mP_n(\mX) \cap (\mP(\mX) - \mC_\delta)}(n+1)\exp(-2n\delta^2) \\
        & \leq & (n+1)^{|\mX| } \exp(-2n\delta^2) \\
        & = & \exp\l(-n \bigg(2\delta^2 - |\mX| \frac{\ln (n+1)}{n}\bigg)\r).
    \end{eqnarray}

Then, regardless of the switcher strategy,
    \begin{equation}
        \E[d(\x^n;\mB)] \leq D + d^\ast \cdot \exp \Bigg(-n \bigg( 2\delta^2 - |\mX| \frac{\ln
        (n+1)}{n}\bigg)\Bigg).
    \end{equation}

So for large $n$ we can get arbitrarily close to distortion $D$ while the rate is at most
$\widetilde{R}(D) + \epsilon$. Using the fact that the IID rate-distortion function is
continuous in $D$ gives us that the coder can achieve at most distortion $D$ on average while
the asymptotic rate is at most $\widetilde{R}(D) + \epsilon$. Since $\epsilon$ is arbitrary,
$R(D) \leq \widetilde{R}(D)$.
\end{proof}

\subsection{Achievability for the switcher} \label{sec:switcherachievability} This section
shows that $R(D) \geq \widetilde{R}(D)$ when the switcher has $1$-step lookahead. We will show
that the switcher can target any distribution $p \in \mC$ and produce a sequence of IID symbols
with distribution $p$. In particular, the switcher can target the distribution that yields
$\max_{p \in \mC} R(p,D)$, so $R(D) \geq \widetilde{R}(D)$.

The switcher will use a memoryless randomized strategy. Let $\mathcal{V} \subseteq \mX$ and
suppose that at some time $k$ the set of symbols available to choose from for the switcher is
exactly $\mathcal{V}$, i.e. $\{x_{1,k}, \ldots, x_{m,k}\} = \mathcal{V}$. Recall
$\beta(\mathcal{V}) \triangleq P(\{x_{1,1}, \ldots, x_{m,1}\} = \mathcal{V})$ is the
probability that at any time the switcher must choose among elements of $\mathcal{V}$ and no other
symbols. Then let $f(x|\mathcal{V})$ be a probability distribution on $\mX$ with support
$\mathcal{V}$, i.e. $f(x|\mathcal{V}) \geq 0,~ \forall ~x \in \mX$, $f(x|\mathcal{V}) = 0$ if
$x \notin \mathcal{V}$, and $\sum_{x\in \mathcal{V}} f(x|\mathcal{V}) = 1$. The switcher will
have such a randomized rule for every nonempty subset $\mathcal{V}$ of $\mX$ such that
$|\mathcal{V}| \leq m$. Let $\mD$ be the set of distributions on $\mX$ that can be achieved
with these kinds of rules,
    \begin{equation}
        \mD = \left\{ \begin{array}{ccc}
        & & p(\cdot) = \sum_{\mathcal{V} \subseteq \mX, |\mathcal{V}|\leq m} \beta(\mathcal{V}) f(\cdot|\mathcal{V}), \\
         p \in \mP(\mX) & : & \forall~ \mathcal{V} \textrm{ s.t. } \mathcal{V} \subseteq \mX, ~ |\mathcal{V}| \leq m, \\
        & & f(\cdot|\mathcal{V}) \textrm{ is a PMF on} ~\mathcal{V} \end{array} \right\}.
    \end{equation}

It is clear by construction that $\mD \subseteq \mC$ because the conditions in
$\mC$ are those that only prevent the switcher from producing symbols that do not occur enough
on average, but put no further restrictions on the switcher. So we need only show that $\mC
\subseteq \mD$. The following gives such a proof by contradiction.

\begin{lemma}[Achievability for switcher]\label{lem:hyperplanearg}
The set relation $\mC \subseteq \mD$ is true.
\end{lemma}\vspace{.3cm}
\begin{proof} Without loss of generality, let $\mX = \{1, \ldots, |\mX|\}$. Suppose $p \in \mC$ but $p \notin \mD$. It is clear that $\mD$
is a convex set. Let us view the probability simplex in $\mathbb{R}^{|\mX|}$. Since $\mD$ is a
convex set, there is a hyperplane through $p$ that does not intersect $\mD$. Hence, there is a
vector $(a_1, \ldots, a_{|\mX|})$ such that $\sum_{i=1}^{|\mX|} a_i p(i) = t$ for some real $t$
but $t < \min_{q \in \mD} \sum_{i=1}^{|\mX|} a_i q(i)$. Without loss of generality, assume $a_1
\geq a_2 \geq \ldots \geq a_{|\mX|}$ (otherwise permute symbols). Now, we will construct
$f(\cdot| \mathcal{V})$ so that the resulting $q$ has $\sum_{i=1}^{|\mX|} a_i p(i) \geq
\sum_{i=1}^{|\mX|} a_i q(i)$, which contradicts the initial assumption. Let
    \begin{equation}
        f(i|\mathcal{V}) \triangleq \left\{ \begin{array}{ccc} 1 & &~ \textrm{ if $i = \max(\mathcal{V})$}\\
        0 & & \textrm{else} \end{array} \right. ,
    \end{equation}
so for example, if $\mathcal{V} = \{1, 5, 6, 9\}$, then $f(9|\mathcal{V}) = 1$ and
$f(i|\mathcal{V}) = 0$ if $i \neq 9$. Call $q$ the distribution on $\mX$ induced by this choice
of $f(\cdot|\mathcal{V})$. Recall that $\kappa(\mV) = P(x_l \in \mV, 1 \leq l \leq m)$. Then,
we have
    \begin{eqnarray}
        \sum_{i=1}^{|\mX|} a_i q(i) & = & a_1 \kappa(\{1\}) + a_2 [\kappa(\{1,2\}) - \kappa(\{1\})] + \nonumber\\
        &  & \hspace{.5in} \cdots + a_{|\mX|} [\kappa(\{1,\ldots, |\mX|\}) - \kappa(\{1,\ldots, |\mX| -
        1\})]
    \end{eqnarray}

By the constraints in the definition (\ref{eqn:defnC}) of $\mC$, we have the following inequalities for $p$:
    \begin{eqnarray}
        p(1) & \geq & \kappa(\{1\}) = q(1) \\
        p(1) + p(2) & \geq & \kappa(\{1,2\}) = q(1) + q(2) \\
        & \vdots & \nonumber\\
        \sum_{i=1}^{|\mX|-1} p(i) & \geq & \kappa(\{1,\ldots, |\mX|-1\}) = \sum_{i=1}^{|\mX| -1} q(i).
    \end{eqnarray}

Therefore, the difference of the objective is
    \begin{eqnarray}
         \sum_{i=1}^{|\mX|} a_i(p(i) - q(i)) & = & a_{|\mX|}\bigg[\sum_{i=1}^{|\mX|} p(i) - q(i)\bigg] + \nonumber\\
         & & (a_{|\mX|-1} - a_{|\mX|})\bigg[\sum_{i=1}^{|\mX|-1} p(i) - q(i)\bigg] + \nonumber\\
        & & \cdots + (a_1-a_2)\bigg[p(1) - q(1)\bigg] \\
        & = & \sum_{i=1}^{|\mX| - 1} (a_i - a_{i+1})\bigg[ \sum_{j=1}^i p(j) - \sum_{j=1}^i q(j)
        \bigg]\\
        & \geq & 0.
    \end{eqnarray}

The last step is true because of the monotonicity in the $a_i$ and the inequalities we derived
earlier. Therefore, we see that $\sum_{i=1}^{|\mX|} a_ip(i) \geq \sum_{i=1}^{|\mX|} a_iq(i)$
for the $p$ we had chosen at the beginning of the proof. This contradicts the assumption that
$\sum_{i=1}^{|\mX|} a_ip(i) < \min_{q \in \mD} \sum_{i=1}^{|\mX|} a_iq(i)$, therefore it must
be that $\mC \subseteq \mD$.
\end{proof}

\section{Proof of Theorem \ref{thm:states}} \label{sec:appendixstates}
It is clear that $R(D) \geq \max_{p \in \mD_{states}} R(p,D)$ because the switcher can select
distributions $f(\cdot|t) \in \overline{\mG}(t)$ for all $t \in \mT$ and upon observing a state
$t$, the switcher can randomly select the switch position according to the convex combination
that yields $f(\cdot|t)$. With this strategy, the AVS is simply an IID source with distribution
$p(\cdot) = \sum_{t} \alpha(t) f(\cdot|t)$. Hence, $R(D) \geq \max_{p \in \mD_{states}}
R(p,D)$.

We will now show that $R(D) \leq \max_{p \in \mD_{states}} R(p,D)$. This can be done in the
same way as in Appendix \ref{sec:appendix1}. We can use the type covering lemma to cover
sequences with types in or very near $\mD_{states}$ and then we need only show that the
probability of $\x^n$ having a type $\epsilon$ far from $\mD_{states}$ goes to $0$ with block
length $n$.

\begin{lemma} Let $p_{\x^n}$ be the type of $\x^n$ and for $\epsilon > 0$ let
$\mD_{{states},\eps}$ be the set of $p \in \mP(\mX)$ with $\mL_1$ distance at most $\epsilon$
from a distribution in $\mD_{states}$. Then, for $\eps > 0$,
    \begin{equation} P(p_{\x^n} \notin \mD_{{states},\epsilon}) \leq 4 |\mT||\mX| \exp(-n\xi(\eps)),\end{equation}
where $\xi(\eps) > 0$ for all $\eps > 0$. So for large $n$, $p_{\x^n}$ is in $\mD_{states,\eps}$
with high probability.
\end{lemma}
\begin{proof} Let $\t^n$ be the $n$-length vector of the observed
states. We assume that the switcher has advance knowledge of all these states before choosing
the switch positions.
    First, we show that with high probability, the states that are observed are strongly
typical. Let $N(t|\t^n)$ be the count of occurrence of $t \in \mT$ in the vector $\t^n$. Fix a
$\delta > 0$ and for $t \in \mT$, define the event
    \begin{equation} A_{\delta}^t = \l\{ \l | \frac{N(t|\t^n)}{n} - \alpha(t)\r| > \delta \r\}. \label{eqn:counts} \end{equation}
Since $N(t|\t^n) = \sum_{i=1}^n {\bf 1}(t_i = t)$ and each term in the sum is an IID Bernoulli
variable with probability of $1$ equal to $\alpha(t)$, we have by Hoeffding's tail inequality
\cite{Hoeffding},
    \begin{equation}  P(A_{\delta}^t) \leq 2 \exp(-2n\delta^2). \end{equation}

Next, we need to show that the substrings output by the AVS at the times when the state is $t$
have a type in or very near $\overline{\mG}(t)$. This will be done by a martingale argument
similar to that given in Lemma 3 of \cite{BergerSourceCodingGame}. Let $\t^\infty$ denote the
infinite state sequence $(t_1,t_2,\ldots)$ and let $\mF_0 = \sigma(\t^\infty)$ be the sigma field
generated by the states $\t^\infty$. For $i = 1, 2, \ldots$, let $\mF_i = \sigma(\t^\infty,
\s^i, \x_1^i, \ldots, \x_m^i)$. Note that $\{\mF_i\}_{i=0}^\infty$ is a filtration and for each
$i$, the $x_i$ is included in $\mF_i$ trivially because $x_i = x_{s_i,i}$.

Let $C_i$ be the $|\mX|$-dimensional unit vector with a $1$ in the position of $x_i$. That is,
$C_i(x) = {\bf 1}(x_i = x)$ for each $x \in \mX$. Define $T_i$ to be
    \begin{equation} T_i = C_i - \E[C_i | \mF_{i-1}] \end{equation}
and let $S_0 = 0$. For $k \geq 1$,
    \begin{equation} S_k = \sum_{i=1}^k T_i. \end{equation}
We claim that $S_k, k \geq 1$ is a martingale\footnote{$S_k$ is a vector, so we show that each
component of the vector is martingale. For ease of notation, we drop the dependence on the
component of the vector until it is explicitly needed.} with respect to the filtration
$\{\mF_i\}$ defined previously. To see this, note that $\E[|S_k|] < \infty$ for all $k$ since
$S_k$ is bounded (not uniformly). Also, $S_k \in \mF_k$ because $T_i \in \mF_i$ for each $i$.
Finally,
    \begin{eqnarray*}
        \E[S_{k+1}|\mF_k] & = & \E[T_{k+1} + S_k|\mF_k] \\
            & = & \E[T_{k+1}|\mF_k] + S_k \\
            & = & \E[C_{k+1} - \E[C_{k+1}|\mF_k]|\mF_k] + S_k \\
            & = & \E[C_{k+1}|\mF_k] - \E[C_{k+1}|\mF_k] + S_k\\
            & = & S_k.
    \end{eqnarray*}
Now, define for each $t \in \mT$,
    \begin{equation} T_i^{t} = T_i \cdot {\bf 1}(t_i = t) \end{equation}
and analogously,
    \begin{equation} S_k^{t} = \sum_{i=1}^k T_i^{t}.
    \end{equation}

It can be easily verified that $S_k^{t}$ is a martingale with respect to $\mF_i$ for each $t
\in \mT$. Expanding, we also see that
    \begin{eqnarray}
        \frac{1}{N(t|\t^n)}S_n^{t} & = & \frac{1}{N(t|\t^n)}\sum_{i=1}^n T_i {\bf 1}(t_i = t) \\
        & = & \frac{1}{N(t|\t^n)}\sum_{i: ~t_i = t} C_i  - \frac{1}{N(t|\t^n)}\sum_{i: ~t_i = t} \E[C_i
        | \mF_{i-1}]. \label{eqn:difference}
    \end{eqnarray}

The first term in the difference above is the type of the output of the AVS during times when
the state is $t$. For any $i$ such that $t_i = t$,
    \begin{equation} \E[C_i|\mF_{i-1}] = \sum_{l=1}^m P(l|\mF_{i-1}) p_l(\cdot|t) \in \overline{\mG}(t).\end{equation}
In the above, $P(l|\mF_{i-1})$ represents the switcher's possibly random strategy because the
switcher chooses the switch position at time $i$ with knowledge of events in $\mF_{i-1}$.  The
source generator's outputs, conditioned on the state at the time are independent of all other
random variables, so $\sum_{l=1}^m P(l|\mF_{i-1}) p_l(\cdot|t)$ is the probability distribution
of the output at time $i$ conditioned on $\mF_{i-1}$.

Thus, the second term in the difference of equation (\ref{eqn:difference}) is in
$\overline{\mG}(t)$ because it is the average of $N(t|\t^n)$ terms in $\overline{\mG}(t)$ and
$\overline{\mG}(t)$ is a convex set. Therefore, $S_n^t/N(t|\t^n)$ measures the difference
between the type of symbols output at times when the state is $t$ and some distribution
guaranteed to be in $\overline{\mG}(t)$.

Let $p_{\x^n}$ be the empirical type of the string $\x^n$, and let $p_{\x^n}^t$ be the
empirical type of the sub-string of $\x^n$ corresponding to the times $i$ when $t_i = t$. Then,
    \begin{eqnarray}
        p_{\x^n} & = & \sum_{t\in \mT} \frac{N(t|\t^n)}{n}
        p_{\x^n}^t.
    \end{eqnarray}

Let $\overline{\mG}(t)_\eps$ be the set of distributions at most $\eps$ in $\mL_1$ distance
from a distribution in $\overline{\mG}(t)$. Recall that for $|\mX|$ dimensional vectors, $\|p -
q\|_\infty < \eps/|\mX|$ implies $\|p-q\|_1 < \epsilon$. Hence, we have
    \begin{eqnarray}
        P\l( \bigcup_{t \in \mT} \l \{ p_{\x^n}^t \notin \overline{\mG}(t)_\eps \r\} \r) & \leq &
        \sum_{t \in \mT} P\l ( \bigcup_{x \in \mX} \l \{ \l | \frac{1}{N(t|\t^n)} S_n^t(x) \r| >
        \frac{\eps}{|\mX|}  \r \} \r)  \\
        & \leq & \sum_{t} \sum_{x} P \l ( \l | \frac{1}{N(t|\t^n)} S_n^t(x) \r| >
        \frac{\eps}{|\mX|}  \r). \label{eqn:everydist} \end{eqnarray}

Let $(A_\delta^t)^c$ denote the complement of the event $A_\delta^t$. So, for every $(t,x)$ we
have
    \begin{eqnarray}
        P\l( \l |\frac{1}{N(t|\t^n)} S_n^t(x)\r| > \frac{\eps}{|\mX|} \r) & \leq &
        P(A_\delta^t) + P\l( \l |\frac{1}{N(t|\t^n)} S_n^t(x)\r| > \frac{\eps}{|\mX|}, (A_\delta^t)^c
        \r) \\
        & \leq & 2\exp(-2n\delta^2) + P\l( \l |\frac{1}{N(t|\t^n)} S_n^t(x)\r| > \frac{\eps}{|\mX|}, (A_\delta^t)^c
        \r).
    \end{eqnarray}
In the event of $(A_\delta^t)^c$, we have $N(t|\t^n) \geq n(\alpha(t) - \delta)$, so
    \begin{eqnarray}
        P\l( \l |\frac{1}{N(t|\t^n)} S_n^t(x)\r| > \frac{\eps}{|\mX|}, (A_\delta^t)^c
        \r) & \leq & P \l ( |S_n^t(x)| > n(\alpha(t) - \delta)\frac{\eps}{|\mX|}, (A_\delta^t)^c
        \r)\\
        & \leq & P \l ( |S_n^t(x)| > n(\alpha(t) - \delta)\frac{\eps}{|\mX|}\r).
    \end{eqnarray}
$S_k^t(x)$ is a martingale with bounded differences since $|S_{k+1}^t(x) - S_k^t(x)| =
|T_{k+1}^t(x)| \leq 1$. Hence, we can apply Azuma's inequality \cite{Azuma} to get
    \begin{eqnarray}
        P \l ( |S_n^t(x)| > n(\alpha(t) - \delta)\frac{\eps}{|\mX|}\r) & \leq & 2 \exp\l ( - n
        \frac{(\alpha(t) - \delta)^2 \eps^2}{2|\mX|^2} \r).
    \end{eqnarray}
Plugging this back into equation (\ref{eqn:everydist}),
    \begin{eqnarray}
        P\l( \bigcup_{t \in \mT} \l \{ p_{\x^n}^t \notin \overline{\mG}(t)_\eps \r\} \r) & \leq
        & 2|\mT||\mX|\l( \exp(-2n\delta^2) + \exp\l ( - n
        \frac{(\alpha_\ast - \delta)^2 \eps^2}{2|\mX|^2} \r)\r) \\
        & \leq & 4|\mX||\mT| \exp(-n \xi(\eps, \delta))
    \end{eqnarray}
where
    \begin{eqnarray} \xi(\eps, \delta) & = & \min \l \{ 2\delta^2, \frac{(\alpha_\ast - \delta)^2
    \eps^2}{2|\mX|^2}\r\} \\
    \alpha_* & \triangleq & \min_{t \in \mT} \alpha(t). \end{eqnarray}
We assume without loss of generality that $\alpha_\ast > 0$ since $\mT$ is finite. We will soon
need that $\delta \leq \eps/|\mT|$, so let
    \begin{equation} \wt{\xi}(\eps) = \max_{0 < \delta < \min\{\eps/|\mT|, \alpha_*\}} \xi(\eps, \delta) \end{equation}
and note that it is always positive provided $\eps > 0$, since $\xi(\eps, \delta) > 0$ whenever
$\delta \in (0,\alpha_\ast)$. Hence,
    \begin{equation} P\l( \bigcup_{t \in \mT} \l \{ p_{\x^n}^t \notin \overline{\mG}(t)_\eps \r\}
    \r)\leq 4|\mX||\mT|\exp(-n\wt{\xi}(\eps)) \stackrel{n}{\longrightarrow} 0.\end{equation}

We have shown that with probability at least $1 - 4|\mX||\mT| \exp(-n\wt{\xi}(\eps))$, for each
$t \in \mT$ there is some $p^t \in \overline{\mG}(t)$ such that $\|p_{\x^n}^t - p^t\|_1 \leq
\epsilon$ and $(A_{\eps/|\mT|}^t)^c$ occurs. Let
    \begin{eqnarray}
        p = \sum_{t \in \mT} \alpha(t) p^t.
    \end{eqnarray}
By construction, $p \in \mD_{states}$. To finish, we show that $\|p_{\x^n} - p\|_1 \leq 2
\eps$.
    \begin{eqnarray}
        \|p_{\x^n} - p \|_1 & = & \sum_{x \in \mX} |p_{\x^n}(x) - p(x)| \\
            & =  & \sum_x \Bigg | \sum_{t \in \mT}
            \frac{N(t|\t^n)}{n}p_{\x^n}^t(x) - \alpha(t) p^t(x)\Bigg| \\
            & \leq & \sum_{t} \sum_x \l| \frac{N(t|\t^n)}{n} p_{\x^n}^t(x) -  \alpha(t)p^t(x) \r |\\
            & = & \sum_{t} \alpha(t)\sum_x \l |
            \frac{N(t|\t^n)}{n\alpha(t)} p_{\x^n}^t(x) - p^t(x) \r| \\
            & \leq & \sum_{t} \alpha(t)\sum_x | p_{\x^n}^t(x) -
              p^t(x) | + \l | \frac{N(t|\t^n)}{n
              \alpha(t)} - 1 \r| p_{\x^n}^t(x).
    \end{eqnarray}

From (\ref{eqn:counts}), we are assumed to be in the event that
    \begin{equation} \l|\frac{N(t|\t^n)}{n\alpha(t)} -
    1 \r| \leq \frac{\delta}{\alpha(t)} \end{equation}
Hence,
    \begin{eqnarray}
        \|p_{\x^n} - p \|_1 & \leq & \sum_{t} \alpha(t) \l (
        \epsilon + \frac{\delta}{\alpha(t)} \r) \\
            & = & \epsilon + |\mT|\delta \leq 2 \epsilon.
    \end{eqnarray}
We have proved $P(p_{\x^n} \notin \mD_{states,2\eps}) \leq 4 |\mX||\mT| \exp(-n\wt{\xi}(\eps))$,
so we arrive at the conclusion of the lemma by letting $\xi(\epsilon) = \wt{\xi}(\epsilon/2)$.

\end{proof}
\section{Proof of Lemma \ref{lem:ratedistortionbound}} \label{sec:appendix2}
 Let $W^\ast_{p,D} = {\textrm{arg min}}_{W \in \mW(p,D)}
I(p,W)$. Then
    \begin{equation} |R(p,D) - R(q,D)| = | I(p, W^\ast_{p,D}) -
    I(q,W^\ast_{q,D}) |. \end{equation}
Consider $d(p,W^\ast_{q,D})$, the distortion of source $p$ across $q$'s distortion $D$
achieving channel.
    \begin{eqnarray}
        d(p, W^\ast_{q,D}) & \leq  & d(q,W^\ast_{q,D}) + |d(p,W^\ast_{q,D}) - d(q,W^\ast_{q,D})| \\
            & = & d(q,W^\ast_{q,D}) + \left | \sum_x \sum_{\hat{x}} (p(x) - q(x)) W^\ast_{q,D}(\hx|x)d(x,\hx) \right| \\
            & \leq & D + \sum_x |p(x) - q(x)| \sum_{\hx} W^\ast_{q,D}(\hx|x)d(x,\hx) \\
            & \leq & D + \|p - q\|_1 d^\ast.
    \end{eqnarray}
By definition, $W^\ast_{q,D}$ is in $\mW(p, d(p,W^\ast_{q,D}))$, so $R(p,d(p,W^\ast_{q,D}))
\leq I(p,W^\ast_{q,D})$.
    \begin{eqnarray}
        R(p,d(p,W^\ast_{q,D})) & \leq & I(p, W^\ast_{q,D}) \\
            & \leq & I(q,W^\ast_{q,D}) + |I(p,W^\ast_{q,D}) -I(q,W^\ast_{q,D})| \\
            & = & R(q,D) + |I(p,W^\ast_{q,D}) -I(q,W^\ast_{q,D})|.
            \label{eqn:midway}
    \end{eqnarray}
Expanding mutual informations yields
    \begin{eqnarray}
    |I(p,W^\ast_{q,D}) - I(q,W^\ast_{q,D})| & = &
            |H(p) + H(pW^\ast_{q,D}) - H(p,W^\ast_{q,D}) \cdots \\
        & & \hspace{4cm} - H(q) - H(qW^\ast_{q,D}) + H(q,W^\ast_{q,D})| \nonumber\\
    & \leq & |H(p) - H(q)| + |H(pW^\ast_{q,D}) - H(qW^\ast_{q,D})| + \cdots \nonumber \\
    & & \hspace{4cm} |H(p,W^\ast_{q,D}) -
    H(q,W^\ast_{q,D})|.
    \end{eqnarray}
Above, for a distribution $p$ on $\mX$ and channel $W$ from $\mX$ to $\mXh$, $H(pW)$ denotes
the entropy of a distribution on $\mXh$ with probabilities $(pW)(\hx) = \sum_x p(x) W(\hx|x)$.
$H(p,W)$ denotes the entropy of the joint source on $\mX \times \mXh$ with probabilities
$(p,W)(x,\hx) = p(x)W(\hx|x)$. It is straightforward to verify that $\|pW - qW\|_1 \leq \|p -
q\|_1$ and $\|(p,W) - (q,W)\|_1 \leq \|p - q\|_1$. So using Lemma \ref{lem:entropybound} three
times, we have
    \begin{eqnarray}
        |I(p,W^\ast_{q, D}) - I(q, W^\ast_{q,D})| & \leq & \|p-q\|_1 \ln
        \frac{|\mX|}{\|p-q\|_1} + \|p-q\|_1 \ln \frac{|\mXh|}{\|p-q\|_1} + \nonumber \\
        & & \hspace{4cm} \|p-q\|_1 \ln \frac{|\mX||\mXh|}{\|p-q\|_1} \\
        & \leq & 3 \|p-q\|_1 \ln \frac{|\mX||\mXh|}{\|p-q\|_1}.
    \end{eqnarray}

Now, we have seen $d(p,W^\ast_{q,D}) \leq D + d^\ast \|p-q\|_1$. We will use the uniform
continuity of $R(p,D)$ in $D$ to bound $| R(p,D) - R(p,D+d^\ast\|p-q\|_1)|$. This will give an
upper bound on $R(p,D) - R(q,D)$ as seen through equation (\ref{eqn:midway}), namely,
    \begin{eqnarray}
        R(p,D) - R(q,D) & \leq & |I(p,W^\ast_{q,D}) -I(q,W^\ast_{q,D})| +
       R(p,D) - R(p,d(p,W^\ast_{q,D})) \\
        & \leq & |I(p,W^\ast_{q,D}) -I(q,W^\ast_{q,D})| +
        R(p,D) - R(p,D+d^\ast \|p-q\|_1),  \label{eqn:midway2}
    \end{eqnarray} where the last step follows because $R(p,D)$ is monotonically decreasing in
    $D$.
For a fixed $p$, the rate-distortion function in $D$ is convex $\cup$ and decreasing and so has
steepest descent at $D = 0$. Therefore, for any $0 \leq D_1, D_2 \leq d^\ast$,
    \begin{equation} |R(p,D_1) - R(p,D_2)| \leq |R(p,0) - R(p,|D_2 - D_1|)|.
    \end{equation}
Hence, we can restrict our attention to continuity of $R(p,D)$ around $D = 0$. By assumption,
$\mW(p,0) \neq \emptyset$ $\forall p \in \mP(\mX)$. Now consider an arbitrary $D > 0$, and let
$W \in \mW(p, D)$. We will show that there is some $W_0 \in \mW(p,0)$ that is close to $W$ in
an $\mL_1$-like sense (relative to the distribution $p$). Since $W \in \mW(p,D)$, we have by
definition
    \begin{eqnarray}
        D & \geq & \sum_x p(x) \sum_{\hx} W(\hx|x)d(x,\hx) \\
         & = & \sum_x p(x) \sum_{\hx: ~d(x,\hx) > 0} W(\hx|x) d(x,\hx) \\
         & \geq & \wt{d} \sum_x p(x) \sum_{\hx: ~d(x,\hx) > 0} W(\hx|x). \label{eqn:distbound}
    \end{eqnarray}
Now, we will construct a channel in $\mW(p,0)$, denoted $W_0$. First, for each $x, \hx$ such
that $d(x,\hx) = 0$, let $V(\hx|x) = W(\hx|x)$. For all other $(x,\hx)$, set $V(\hx|x) = 0$.
Note that $V$ is not a channel matrix if $W \notin \mW(p,0)$ since it is missing some
probability mass. To create $W_0$, for each $x$, we redistribute the missing mass from
$V(\cdot|x)$ to the pairs $(x,\hx)$ with $d(x,\hx) = 0$. Namely, for $(x,\hx)$ with $d(x,\hx) =
0$, we define
    \begin{eqnarray}
        W_0(\hx|x) & = & V(\hx|x) + \frac{\sum_{\hat{x}':~d(x,\hat{x}') > 0} W(\hat{x}'|x)}{|\{\hat{x}':~d(x,\hat{x}')
        = 0\}|}.
    \end{eqnarray}
For all $(x,\hx)$ with $d(x,\hx) > 0$, define $W_0(\hx|x) = 0$. So, $W_0$ is a valid channel in
$\mW(p,0)$. Now for a fixed $x \in \mX$,
    \begin{eqnarray}
        \sum_{\hx} |W(\hx|x) - W_0(\hx|x)| & = & \sum_{\hx:~d(x,\hx) > 0} W(\hx|x) +
        \sum_{\hx:~d(x, \hx) = 0} |W(\hx|x) - W_0(\hx|x)| \\
            & = & \sum_{\hx:~d(x,\hx) > 0} W(\hx|x) + \cdots \\
            & & \hspace{2cm}
                            \sum_{\hx:~d(x, \hx) = 0} \left | W(\hx|x) - W(\hx|x) - \frac{\sum_{\hat{x}':~d(x,\hat{x}') > 0} W(\hat{x}'|x)}{|\{\hat{x}':~d(x,\hat{x}')
                            = 0\}|} \right| \nonumber \\
            & = & 2\sum_{\hx:~d(x,\hx) > 0} W(\hx|x).
    \end{eqnarray}
Therefore, using (\ref{eqn:distbound})
    \begin{eqnarray}
        \sum_x p(x) \sum_{\hx} |W(\hx|x) - W_0(\hx|x)| & \leq & \frac{2D}{\wt{d}}.
    \end{eqnarray}
So, for $W = W^\ast_{p,D}$, there is a $W_0 \in \mW(p,0)$ with the above `modified $\mL_1$
distance' with respect to $p$ between $W$ and $W_0$ being less than $2D / \wt{d}$. Going back
to the bound on $|R(p,0) - R(p,D)|$,
    \begin{eqnarray}
        |R(p,0) - R(p,D)| & = & \min_{W \in \mW(p,0)} I(p,W) - I(p,W^\ast_{p,D}) \\
            & \leq & I(p,W_0) - I(p,W^\ast_{p,D}) \\
            & \leq & |H(pW_0) - H(pW^\ast_{p,D})| + |H(p,W_0) - H(p,W^\ast_{p,D})|.
    \end{eqnarray}
Now, note that the $\mL_1$ distance between $pW_0$ and $pW^\ast_{p,D}$ is
    \begin{eqnarray}
        \|pW_0 - pW^\ast_{p,D}\|_1 & = & \sum_{\hx} \left |\sum_x p(x)W_0(\hx|x) - p(x)W^\ast_{p,D}(\hx|x) \right | \\
        & \leq & \sum_x p(x) \sum_{\hx} |W_0(\hx|x) - W^\ast_{p,D}(\hx|x)| \\
        & \leq & \frac{2D}{\wt{d}}.
    \end{eqnarray}
Similarly, $\| (p,W_0) - (p,W^\ast_{p,D})\|_1 \leq 2D/\wt{d}$.

Now, assuming $D \leq \wt{d}/4$, we can again invoke Lemma \ref{lem:entropybound} to get
    \begin{eqnarray}
        |R(p,0) - R(p,D)| & \leq & \frac{2D}{\wt{d}} \ln \frac{\wt{d}
        |\mX|}{2D} + \frac{2D}{\wt{d}} \ln \frac{\wt{d}
        |\mX||\mXh|}{2D} \\
        & \leq & \frac{4D}{\wt{d}} \ln \frac{\wt{d}|\mX||\mXh|}{2D}. \label{eqn:distcontinuity}\end{eqnarray}

Going back to (\ref{eqn:midway2}), we see that if $\|p - q\|_1 \leq
\frac{\wt{d}}{4d^\ast}$,
    \begin{eqnarray}
        |R(p,d+d^\ast\|p-q\|_1)) - R(p,D)| & \leq & \frac{4 d^\ast \|p -
        q\|_1}{\wt{d}} \ln \frac{\wt{d} |\mX||\mXh|}{2d^\ast \|p - q\|_1} \\
            & \leq & \frac{4 d^\ast \|p - q\|_1}{\wt{d}} \ln \frac{ |\mX||\mXh|}{\|p - q\|_1}.
    \end{eqnarray}
The last step follows because $\wt{d}/d^\ast \leq 1$. Substituting into equation
(\ref{eqn:midway2}) gives
    \begin{eqnarray}
        R(p,D) - R(q,D) & \leq & 3 \|p - q\|_1 \ln \frac{|\mX||\mXh|}{\|p - q\|_1} +
        4 \frac{d^\ast}{\wt{d}} \|p -q\|_1 \ln \frac{|\mX||\mXh|}{\|p-q\|_1} \\
        & \leq & \frac{7d^\ast}{\wt{d}} \|p -q\|_1 \ln \frac{|\mX||\mXh|}{\|p-q\|_1}. \end{eqnarray}
Finally, this bound holds uniformly on $p$ and $q$ as long as the condition on $\|p - q\|_1$ is
satisfied. Therefore, we can interchange $p$ and $q$ to get the other side of the inequality.
    \begin{eqnarray}
        R(q,D) - R(p,D) & \leq & \frac{7d^\ast}{\wt{d}} \|p -q\|_1 \ln
        \frac{|\mX||\mXh|}{\|p-q\|_1}.
    \end{eqnarray}
This concludes the proof.

\section{Proof of Lemma \ref{lem:ratedistortionbound2}} \label{sec:appendix3}
We now assume $d:\mX \times\mXh \to [0,d^\ast]$ to be arbitrary. However, we let
    \begin{equation} d_0(x,\hx) = d(x,\hx) - \min_{\wt{x} \in \mXh} d(x,\wt{x}) \end{equation}
so that Lemma \ref{lem:ratedistortionbound} applies to $d_0$.  Let $R_0(p,D)$ be the IID
rate-distortion function for $p \in \mP(\mX)$ at distortion $D$ with respect to distortion
measure $d_0(x,\hx)$. By definition, $R(p,D)$ is the IID rate-distortion function for $p$ with
respect to distortion measure $d(x,\hx)$. From Problem 13.4 of \cite{CoverBook}, for any $D
\geq D_{\min} (p)$,
    \begin{equation} R(p,D) = R_0(p, D - D_{\min}(p)). \end{equation}
Hence, for $p,q \in \mP(\mX)$, $D \geq \max(D_{\min}(p), D_{\min}(q))$,
    \begin{eqnarray} |R(p,D) - R(q,D)| & = & |R_0(p,D - D_{\min}(p)) - R_0(q,D - D_{\min}(q)|\\
        & \leq & |R_0(p,D - D_{\min}(p)) - R_0(p,D-D_{\min}(q))| + \nonumber \\
            & & \hspace{3cm}|R_0(p, D-D_{\min}(q)) - R_0(q,D - D_{\min}(q))|. \label{eqn:twoterms}\end{eqnarray}
Now, we note that $|D_{\min}(p) - D_{\min}(q)| \leq d^\ast \|p-q\|_1$. The first term of
equation (\ref{eqn:twoterms}) can be bounded using equation (\ref{eqn:distcontinuity}) and the
second term of (\ref{eqn:twoterms}) can be bounded using Lemma \ref{lem:ratedistortionbound}.
The first term can be bounded if $\|p-q\|_1 \leq \wt{d}_0/4d^\ast$ and the second can be
bounded if $\|p-q\|_1 \leq \wt{d}_0 /4 d^\ast_0$. Since $d_0^\ast \leq d^\ast$, we only require
$\|p-q\|_1 \leq \wt{d}_0/4 d^\ast$.
    \begin{eqnarray}
        |R(p,D) - R(q,D)| & \leq & \frac{4d^\ast}{\wt{d}_0}\|p-q\|_1\ln \frac{\wt{d}_0
        |\mX||\mXh|}{2 d^\ast \|p-q\|_1} + \frac{7d_0^\ast}{\wt{d}_0} \|p-q\|_1 \ln
        \frac{|\mX||\mXh|}{\|p-q\|_1} \\
        & \leq &  \frac{4d^\ast}{\wt{d}_0}\|p-q\|_1\ln \frac{
        |\mX||\mXh|}{\|p-q\|_1} + \frac{7d^\ast}{\wt{d}_0} \|p-q\|_1 \ln
        \frac{|\mX||\mXh|}{\|p-q\|_1}.\end{eqnarray}

\end{document}

%% file: switcher.pstex_t
\begin{picture}(0,0)%
\includegraphics{switcher.pstex}%
\end{picture}%
\setlength{\unitlength}{2763sp}%
\begingroup\makeatletter\ifx\SetFigFont\undefined%
\gdef\SetFigFont#1#2#3#4#5{%
  \reset@font\fontsize{#1}{#2pt}%
  \fontfamily{#3}\fontseries{#4}\fontshape{#5}%
  \selectfont}%
\fi\endgroup%
\begin{picture}(3844,2856)(1154,-2683)
\put(1701,-1674){\makebox(0,0)[lb]{\smash{{\SetFigFont{14}{16.8}{\rmdefault}{\mddefault}{\updefault}$\vdots$}}}}
\put(1714,-992){\makebox(0,0)[lb]{\smash{{\SetFigFont{11}{13.2}{\rmdefault}{\mddefault}{\updefault}$p_2$}}}}
\put(1714,-361){\makebox(0,0)[lb]{\smash{{\SetFigFont{11}{13.2}{\rmdefault}{\mddefault}{\updefault}$p_1$}}}}
\put(1714,-2299){\makebox(0,0)[lb]{\smash{{\SetFigFont{11}{13.2}{\rmdefault}{\mddefault}{\updefault}$p_m$}}}}
\put(4176,-1174){\makebox(0,0)[lb]{\smash{{\SetFigFont{11}{13.2}{\rmdefault}{\mddefault}{\updefault}$(x_1, x_2, \ldots)$}}}}
\end{picture}%

%% file: switcherwithstate.pstex_t
\begin{picture}(0,0)%
\includegraphics{switcherwithstate.pstex}%
\end{picture}%
\setlength{\unitlength}{3000sp}%
\begingroup\makeatletter\ifx\SetFigFont\undefined%
\gdef\SetFigFont#1#2#3#4#5{%
  \reset@font\fontsize{#1}{#2pt}%
  \fontfamily{#3}\fontseries{#4}\fontshape{#5}%
  \selectfont}%
\fi\endgroup%
\begin{picture}(8241,4084)(50,-3459)
\put(7584,-673){\makebox(0,0)[lb]{\smash{{\SetFigFont{9}{10.8}{\rmdefault}{\mddefault}{\updefault}$x_1, x_2, x_3, \ldots$}}}}
\put(4466,-1689){\makebox(0,0)[lb]{\smash{{\SetFigFont{9}{10.8}{\rmdefault}{\mddefault}{\updefault}$x_{m,1}, x_{m,2}, \ldots$}}}}
\put(4442,201){\makebox(0,0)[lb]{\smash{{\SetFigFont{9}{10.8}{\rmdefault}{\mddefault}{\updefault}$x_{1,1}, x_{1,2}, \ldots$}}}}
\put(4418,-507){\makebox(0,0)[lb]{\smash{{\SetFigFont{9}{10.8}{\rmdefault}{\mddefault}{\updefault}$x_{2,1}, x_{2,2}, \ldots$}}}}
\put(1714,-755){\makebox(0,0)[lb]{\smash{{\SetFigFont{9}{10.8}{\rmdefault}{\mddefault}{\updefault}$t_1,t_2,..$}}}}
\put(710,-578){\makebox(0,0)[lb]{\smash{{\SetFigFont{11}{13.2}{\rmdefault}{\mddefault}{\updefault}State}}}}
\put(473,-1110){\makebox(0,0)[lb]{\smash{{\SetFigFont{11}{13.2}{\rmdefault}{\mddefault}{\updefault}Generator}}}}
\put(2954,-342){\makebox(0,0)[lb]{\smash{{\SetFigFont{11}{13.2}{\rmdefault}{\mddefault}{\updefault}Symbol}}}}
\put(2836,-755){\makebox(0,0)[lb]{\smash{{\SetFigFont{11}{13.2}{\rmdefault}{\mddefault}{\updefault}Generator}}}}
\put(2658,-1110){\makebox(0,0)[lb]{\smash{{\SetFigFont{9}{10.8}{\rmdefault}{\mddefault}{\updefault}$p(x_1,\ldots,x_m|t)$}}}}
\end{picture}%

%% file: example.pstex_t
\begin{picture}(0,0)%
\includegraphics{example.pstex}%
\end{picture}%
\setlength{\unitlength}{2960sp}%
\begingroup\makeatletter\ifx\SetFigFont\undefined%
\gdef\SetFigFont#1#2#3#4#5{%
  \reset@font\fontsize{#1}{#2pt}%
  \fontfamily{#3}\fontseries{#4}\fontshape{#5}%
  \selectfont}%
\fi\endgroup%
\begin{picture}(6895,3706)(451,-3317)
\put(6804,-720){\makebox(0,0)[lb]{\smash{{\SetFigFont{9}{10.8}{\rmdefault}{\mddefault}{\updefault}$x_1,x_2,\ldots$}}}}
\put(3143,-153){\makebox(0,0)[lb]{\smash{{\SetFigFont{9}{10.8}{\rmdefault}{\mddefault}{\updefault}$x_{1,1}, x_{1,2},x_{1,3}, \ldots$}}}}
\put(3095,-1334){\makebox(0,0)[lb]{\smash{{\SetFigFont{9}{10.8}{\rmdefault}{\mddefault}{\updefault}$x_{2,1}, x_{2,2},x_{2,3}, \ldots$}}}}
\put(5316,-2444){\makebox(0,0)[lb]{\smash{{\SetFigFont{9}{10.8}{\rmdefault}{\mddefault}{\updefault}$s_1,s_2,\ldots$}}}}
\put(4324,-2114){\makebox(0,0)[lb]{\smash{{\SetFigFont{11}{13.2}{\rmdefault}{\mddefault}{\updefault}Switch}}}}
\put(4206,-2468){\makebox(0,0)[lb]{\smash{{\SetFigFont{11}{13.2}{\rmdefault}{\mddefault}{\updefault}Selection}}}}
\put(804,-295){\makebox(0,0)[lb]{\smash{{\SetFigFont{11}{13.2}{\rmdefault}{\mddefault}{\updefault}${\mathcal{B}}(1/4)$}}}}
\put(804,-1500){\makebox(0,0)[lb]{\smash{{\SetFigFont{11}{13.2}{\rmdefault}{\mddefault}{\updefault}${\mathcal{B}}(1/3)$}}}}
\end{picture}%

%% file: example_cheating.pstex_t
\begin{picture}(0,0)%
\includegraphics{example_cheating.pstex}%
\end{picture}%
\setlength{\unitlength}{2960sp}%
\begingroup\makeatletter\ifx\SetFigFont\undefined%
\gdef\SetFigFont#1#2#3#4#5{%
  \reset@font\fontsize{#1}{#2pt}%
  \fontfamily{#3}\fontseries{#4}\fontshape{#5}%
  \selectfont}%
\fi\endgroup%
\begin{picture}(6895,3352)(451,-2963)
\put(6804,-720){\makebox(0,0)[lb]{\smash{{\SetFigFont{9}{10.8}{\rmdefault}{\mddefault}{\updefault}$x_1,x_2,\ldots$}}}}
\put(3143,-153){\makebox(0,0)[lb]{\smash{{\SetFigFont{9}{10.8}{\rmdefault}{\mddefault}{\updefault}$x_{1,1}, x_{1,2},x_{1,3}, \ldots$}}}}
\put(3095,-1334){\makebox(0,0)[lb]{\smash{{\SetFigFont{9}{10.8}{\rmdefault}{\mddefault}{\updefault}$x_{2,1}, x_{2,2},x_{2,3}, \ldots$}}}}
\put(5316,-2444){\makebox(0,0)[lb]{\smash{{\SetFigFont{9}{10.8}{\rmdefault}{\mddefault}{\updefault}$s_1,s_2,\ldots$}}}}
\put(4324,-2114){\makebox(0,0)[lb]{\smash{{\SetFigFont{11}{13.2}{\rmdefault}{\mddefault}{\updefault}Switch}}}}
\put(4206,-2468){\makebox(0,0)[lb]{\smash{{\SetFigFont{11}{13.2}{\rmdefault}{\mddefault}{\updefault}Selection}}}}
\put(804,-295){\makebox(0,0)[lb]{\smash{{\SetFigFont{11}{13.2}{\rmdefault}{\mddefault}{\updefault}${\mathcal{B}}(1/4)$}}}}
\put(804,-1500){\makebox(0,0)[lb]{\smash{{\SetFigFont{11}{13.2}{\rmdefault}{\mddefault}{\updefault}${\mathcal{B}}(1/3)$}}}}
\end{picture}%

%% file: example_line_simplex.pstex_t
\begin{picture}(0,0)%
\includegraphics{example_line_simplex.pstex}%
\end{picture}%
\setlength{\unitlength}{2960sp}%
\begingroup\makeatletter\ifx\SetFigFont\undefined%
\gdef\SetFigFont#1#2#3#4#5{%
  \reset@font\fontsize{#1}{#2pt}%
  \fontfamily{#3}\fontseries{#4}\fontshape{#5}%
  \selectfont}%
\fi\endgroup%
\begin{picture}(7250,1727)(804,-1689)
\put(6851,-909){\makebox(0,0)[lb]{\smash{{\SetFigFont{11}{13.2}{\rmdefault}{\mddefault}{\updefault}$P(x = 1)$}}}}
\put(2080,-531){\makebox(0,0)[lb]{\smash{{\SetFigFont{11}{13.2}{\rmdefault}{\mddefault}{\updefault}$1/4$}}}}
\put(1277,-1287){\makebox(0,0)[lb]{\smash{{\SetFigFont{11}{13.2}{\rmdefault}{\mddefault}{\updefault}$1/12$}}}}
\put(2788,-531){\makebox(0,0)[lb]{\smash{{\SetFigFont{11}{13.2}{\rmdefault}{\mddefault}{\updefault}$1/3$}}}}
\put(3828,-1287){\makebox(0,0)[lb]{\smash{{\SetFigFont{11}{13.2}{\rmdefault}{\mddefault}{\updefault}$1/2$}}}}
\put(2599,-153){\makebox(0,0)[lb]{\smash{{\SetFigFont{11}{13.2}{\rmdefault}{\mddefault}{\updefault}$\conv(\mG)$}}}}
\put(2505,-1618){\makebox(0,0)[lb]{\smash{{\SetFigFont{11}{13.2}{\rmdefault}{\mddefault}{\updefault}$\mC$}}}}
\put(804,-531){\makebox(0,0)[lb]{\smash{{\SetFigFont{11}{13.2}{\rmdefault}{\mddefault}{\updefault}$0$}}}}
\put(6615,-531){\makebox(0,0)[lb]{\smash{{\SetFigFont{11}{13.2}{\rmdefault}{\mddefault}{\updefault}$1$}}}}
\end{picture}%

%% file: example_mod2.pstex_t
\begin{picture}(0,0)%
\includegraphics{example_mod2.pstex}%
\end{picture}%
\setlength{\unitlength}{2960sp}%
\begingroup\makeatletter\ifx\SetFigFont\undefined%
\gdef\SetFigFont#1#2#3#4#5{%
  \reset@font\fontsize{#1}{#2pt}%
  \fontfamily{#3}\fontseries{#4}\fontshape{#5}%
  \selectfont}%
\fi\endgroup%
\begin{picture}(6895,3352)(451,-2963)
\put(3143,-153){\makebox(0,0)[lb]{\smash{{\SetFigFont{9}{10.8}{\rmdefault}{\mddefault}{\updefault}$x_{1,1}, x_{1,2},x_{1,3}, \ldots$}}}}
\put(3095,-1334){\makebox(0,0)[lb]{\smash{{\SetFigFont{9}{10.8}{\rmdefault}{\mddefault}{\updefault}$x_{2,1}, x_{2,2},x_{2,3}, \ldots$}}}}
\put(6804,-720){\makebox(0,0)[lb]{\smash{{\SetFigFont{9}{10.8}{\rmdefault}{\mddefault}{\updefault}$x_1,x_2,\ldots$}}}}
\put(5316,-2444){\makebox(0,0)[lb]{\smash{{\SetFigFont{9}{10.8}{\rmdefault}{\mddefault}{\updefault}$s_1,s_2,\ldots$}}}}
\put(3072,-2173){\makebox(0,0)[lb]{\smash{{\SetFigFont{9}{10.8}{\rmdefault}{\mddefault}{\updefault}$t_1,t_2,\ldots$}}}}
\put(4324,-2114){\makebox(0,0)[lb]{\smash{{\SetFigFont{11}{13.2}{\rmdefault}{\mddefault}{\updefault}Switch}}}}
\put(4206,-2468){\makebox(0,0)[lb]{\smash{{\SetFigFont{11}{13.2}{\rmdefault}{\mddefault}{\updefault}Selection}}}}
\put(804,-295){\makebox(0,0)[lb]{\smash{{\SetFigFont{11}{13.2}{\rmdefault}{\mddefault}{\updefault}${\mathcal{B}}(1/4)$}}}}
\put(804,-1500){\makebox(0,0)[lb]{\smash{{\SetFigFont{11}{13.2}{\rmdefault}{\mddefault}{\updefault}${\mathcal{B}}(1/3)$}}}}
\end{picture}%

%% file: example_x2.pstex_t
\begin{picture}(0,0)%
\includegraphics{example_x2.pstex}%
\end{picture}%
\setlength{\unitlength}{2960sp}%
\begingroup\makeatletter\ifx\SetFigFont\undefined%
\gdef\SetFigFont#1#2#3#4#5{%
  \reset@font\fontsize{#1}{#2pt}%
  \fontfamily{#3}\fontseries{#4}\fontshape{#5}%
  \selectfont}%
\fi\endgroup%
\begin{picture}(6895,3352)(451,-2963)
\put(6804,-720){\makebox(0,0)[lb]{\smash{{\SetFigFont{9}{10.8}{\rmdefault}{\mddefault}{\updefault}$x_1,x_2,\ldots$}}}}
\put(3143,-153){\makebox(0,0)[lb]{\smash{{\SetFigFont{9}{10.8}{\rmdefault}{\mddefault}{\updefault}$x_{1,1}, x_{1,2},x_{1,3}, \ldots$}}}}
\put(3095,-1334){\makebox(0,0)[lb]{\smash{{\SetFigFont{9}{10.8}{\rmdefault}{\mddefault}{\updefault}$x_{2,1}, x_{2,2},x_{2,3}, \ldots$}}}}
\put(5316,-2444){\makebox(0,0)[lb]{\smash{{\SetFigFont{9}{10.8}{\rmdefault}{\mddefault}{\updefault}$s_1,s_2,\ldots$}}}}
\put(4324,-2114){\makebox(0,0)[lb]{\smash{{\SetFigFont{11}{13.2}{\rmdefault}{\mddefault}{\updefault}Switch}}}}
\put(4206,-2468){\makebox(0,0)[lb]{\smash{{\SetFigFont{11}{13.2}{\rmdefault}{\mddefault}{\updefault}Selection}}}}
\put(804,-295){\makebox(0,0)[lb]{\smash{{\SetFigFont{11}{13.2}{\rmdefault}{\mddefault}{\updefault}${\mathcal{B}}(1/4)$}}}}
\put(804,-1500){\makebox(0,0)[lb]{\smash{{\SetFigFont{11}{13.2}{\rmdefault}{\mddefault}{\updefault}${\mathcal{B}}(1/3)$}}}}
\put(2658,-2173){\makebox(0,0)[lb]{\smash{{\SetFigFont{9}{10.8}{\rmdefault}{\mddefault}{\updefault}$t_1,t_2,\ldots$}}}}
\end{picture}%

%% file: example_bsc.pstex_t
\begin{picture}(0,0)%
\includegraphics{example_bsc.pstex}%
\end{picture}%
\setlength{\unitlength}{2960sp}%
\begingroup\makeatletter\ifx\SetFigFont\undefined%
\gdef\SetFigFont#1#2#3#4#5{%
  \reset@font\fontsize{#1}{#2pt}%
  \fontfamily{#3}\fontseries{#4}\fontshape{#5}%
  \selectfont}%
\fi\endgroup%
\begin{picture}(6895,3352)(451,-2963)
\put(6804,-720){\makebox(0,0)[lb]{\smash{{\SetFigFont{9}{10.8}{\rmdefault}{\mddefault}{\updefault}$x_1,x_2,\ldots$}}}}
\put(3143,-153){\makebox(0,0)[lb]{\smash{{\SetFigFont{9}{10.8}{\rmdefault}{\mddefault}{\updefault}$x_{1,1}, x_{1,2},x_{1,3}, \ldots$}}}}
\put(3095,-1334){\makebox(0,0)[lb]{\smash{{\SetFigFont{9}{10.8}{\rmdefault}{\mddefault}{\updefault}$x_{2,1}, x_{2,2},x_{2,3}, \ldots$}}}}
\put(5316,-2444){\makebox(0,0)[lb]{\smash{{\SetFigFont{9}{10.8}{\rmdefault}{\mddefault}{\updefault}$s_1,s_2,\ldots$}}}}
\put(4324,-2114){\makebox(0,0)[lb]{\smash{{\SetFigFont{11}{13.2}{\rmdefault}{\mddefault}{\updefault}Switch}}}}
\put(4206,-2468){\makebox(0,0)[lb]{\smash{{\SetFigFont{11}{13.2}{\rmdefault}{\mddefault}{\updefault}Selection}}}}
\put(804,-295){\makebox(0,0)[lb]{\smash{{\SetFigFont{11}{13.2}{\rmdefault}{\mddefault}{\updefault}${\mathcal{B}}(1/4)$}}}}
\put(804,-1500){\makebox(0,0)[lb]{\smash{{\SetFigFont{11}{13.2}{\rmdefault}{\mddefault}{\updefault}${\mathcal{B}}(1/3)$}}}}
\put(3131,-2114){\makebox(0,0)[lb]{\smash{{\SetFigFont{9}{10.8}{\rmdefault}{\mddefault}{\updefault}$t_1,t_2,\ldots$}}}}
\put(2269,-2326){\makebox(0,0)[lb]{\smash{{\SetFigFont{9}{10.8}{\rmdefault}{\mddefault}{\updefault}$BSC(\delta)$}}}}
\end{picture}%

%% file: cheating_switcher_paper_merged_arxiv.bbl
\begin{thebibliography}{10}
\providecommand{\url}[1]{#1} \csname url@rmstyle\endcsname \providecommand{\newblock}{\relax}
\providecommand{\bibinfo}[2]{#2}
\providecommand\BIBentrySTDinterwordspacing{\spaceskip=0pt\relax}
\providecommand\BIBentryALTinterwordstretchfactor{4}
\providecommand\BIBentryALTinterwordspacing{\spaceskip=\fontdimen2\font plus
\BIBentryALTinterwordstretchfactor\fontdimen3\font minus
  \fontdimen4\font\relax}
\providecommand\BIBforeignlanguage[2]{{%
\expandafter\ifx\csname l@#1\endcsname\relax
\typeout{** WARNING: IEEEtran.bst: No hyphenation pattern has been}%
\typeout{** loaded for the language `#1'. Using the pattern for}%
\typeout{** the default language instead.}%
\else \language=\csname l@#1\endcsname \fi #2}}

\bibitem{ISIT07}
H.~Palaiyanur, C.~Chang, and A.~Sahai, ``The source coding game with a cheating
  switcher,'' in \emph{Proc. Int. Symp. Inform. Theory}, Nice, France, June
  2007.

\bibitem{BajcsyVision}
R.~Bajcsy, ``Active perception,'' \emph{Proceedings of the IEEE}, vol.~76,
  no.~8, pp. 966--1005, Aug. 1988.

\bibitem{VetterliPlenoptic}
A.~Chebira, P.~Dragotti, L.~Sbaiz, and M.~Vetterli, ``Sampling and
  interpolation of the plenoptic function,'' in \emph{Proc.of IEEE
  International Conference on Image Processing}, Barcelona, Spain, Sept. 2003.

\bibitem{longmanWashingtonMonthly}
P.~Longman, ``The best care anywhere,'' \emph{{W}ashington {Monthly}}, Jan.
  2005.

\bibitem{BergerSourceCodingGame}
T.~Berger, ``The source coding game,'' \emph{{IEEE} Transactions on Information
  Theory}, vol.~17, pp. 71--76, Jan. 1971.

\bibitem{ShannonCausalSI}
C.~Shannon, ``Channels with side information at the transmitter,'' \emph{IBM J.
  Res. Devel.}, vol.~2, pp. 289--293, Oct. 1958.

\bibitem{GelfandPinsker}
S.~Gelfand and M.~Pinsker, ``Coding for channel with random parameters,''
  \emph{Probl. Pered. Inform. (Probl. Inf. Transm.)}, vol.~9, pp. 19--31, 1980.

\bibitem{CostaDirtyPaperCoding}
M.~H. Costa, ``Writing on dirty paper,'' \emph{{IEEE} Transactions on
  Information Theory}, vol.~29, pp. 439--441, May 1983.

\bibitem{WillemsDirtyTape1}
F.~Willems, ``On {G}aussian channels with side information at the
  transmitter,'' in \emph{Proc. Int. Symp. Inform. Theory}, Benelux, Enschede,
  The Netherlands, May 1988, pp. 129--135.

\bibitem{WillemsDirtyTape2}
------, ``Signalling for the {G}aussian channel with side information at the
  transmitter,'' in \emph{Proc. Int. Symp. Inform. Theory}, Sorrento, Italy,
  June 2000.

\bibitem{ErezLatticesInterference}
U.~Erez, S.~S. (Shitz), and R.~Zamir, ``Capacity and lattice strategies for
  canceling known interference,'' \emph{{IEEE} Transactions on Information
  Theory}, vol.~51, Nov. 2005.

\bibitem{Agarwal}
\BIBentryALTinterwordspacing M.~Agarwal, A.~Sahai, and S.~Mitter, ``Coding into a source: a
direct inverse
  rate-distortion theorem,'' in \emph{Forty-fourth Allerton Conference on
  Communication, Control, and Computing}, Monticello, IL, Sept. 2006. [Online].
  Available: \url{http://arxiv.org/abs/cs.IT/0610142}
\BIBentrySTDinterwordspacing

\bibitem{NeuhoffGilbert}
D.~Neuhoff and R.~K. Gilbert, ``Causal source codes,'' \emph{{IEEE}
  Transactions on Information Theory}, vol.~28, pp. 701--713, Sept. 1982.

\bibitem{WeissmanMerhavSI}
T.~Weissman and N.~Merhav, ``On causal source codes with side information,''
  \emph{{IEEE} Transactions on Information Theory}, vol.~51, pp. 4003--4013,
  Nov. 2005.

\bibitem{TatikondaSRD}
S.~Tatikonda, A.~Sahai, and S.~Mitter, ``Stochastic linear control over a
  communication channel,'' \emph{{IEEE} Transactions on Automatic Control},
  vol.~49, no.~9, pp. 1549--1561, Sept. 2004.

\bibitem{ShannonRateDistortion}
C.~Shannon, ``Coding theorems for a discrete source with a fidelity
  criterion,'' in \emph{IRE Natl. Conv. Rec.}, 1959, pp. 142--163.

\bibitem{WolfowitzRateDistortion}
J.~Wolfowitz, ``Approximation with a fidelity criterion,'' in \emph{5th
  Berkeley Symp. on Math. Stat. and Prob.}, vol.~1.\hskip 1em plus 0.5em minus
  0.4em\relax Berkeley, California: University of California, Press, 1967, pp.
  565--573.

\bibitem{SakrisonCompoundSources}
D.~Sakrison, ``The rate-distortion function for a class of sources,''
  \emph{Information and Control}, vol.~15, pp. 165--195, Mar. 1969.

\bibitem{CsiszarBook}
I.~Csiszar and J.~Korner, \emph{Information Theory: Coding Theorems for
  Discrete Memoryless Systems}, 2nd~ed.\hskip 1em plus 0.5em minus 0.4em\relax
  New York, NY: Academic Press, 1997.

\bibitem{CoverBook}
T.~Cover and J.~Thomas, \emph{Elements of Information Theory}.\hskip 1em plus
  0.5em minus 0.4em\relax New York, NY: John Wiley and Sons, 1991.

\bibitem{GallagerBook}
R.~Gallager, \emph{Information Theory and Reliable Communication}.\hskip 1em
  plus 0.5em minus 0.4em\relax New York,NY: John Wiley and Sons, 1971.

\bibitem{AhlswedeExtremalProperties}
R.~Ahlswede, ``Extremal properties of rate-distortion functions,'' \emph{{IEEE}
  Transactions on Information Theory}, vol.~36, pp. 166--171, Jan. 1990.

\bibitem{HarrisonEstimationRateDistortion}
\BIBentryALTinterwordspacing M.~Harrison and I.~Kontoyiannis, ``Estimation of the
rate-distortion
  function,'' 2007. [Online]. Available:
  \url{http://arxiv.org/abs/cs/0702018v1}
\BIBentrySTDinterwordspacing

\bibitem{L1Deviation}
\BIBentryALTinterwordspacing T.~Weissman, E.~Ordentlich, G.~Seroussi, S.~Verdu, and M.~L.
Weinberger,
  ``Inequalities for the $l_1$ deviation of the empirical distribution,''
  Hewlett-Packard Labs, Tech. Rep., 2003. [Online]. Available:
  \url{http://www.hpl.hp.com/techreports/2003/HPL-2003-97R1.html}
\BIBentrySTDinterwordspacing

\bibitem{DobrushinMemory}
R.~Dobrushin, ``Unified methods for the transmission of information: The
  general case,'' \emph{Sov. Math.}, vol.~4, pp. 284--292, 1963.

\bibitem{Hoeffding}
W.~Hoeffding, ``Probability inequalities for sums of bounded random
  variables,'' \emph{Journal of the American Statistical Association}, vol.~58,
  no. 301, pp. 13--30, Mar 1963.

\bibitem{Azuma}
K.~Azuma, ``Weighted sums of certain dependent random variables,'' \emph{Tohoku
  Math. Journal}, vol.~19, pp. 357 -- 367, 1967.

\end{thebibliography}
